# Macroscale vortex impingement at a porous-fluid interface induces local heat-transfer enhancement

**Thibaut K. Kemayo, Vishal Srikanth, Justin Courter, Rodrigo R. Caballero, and Andrey V. Kuznetsov***

Department of Mechanical and Aerospace Engineering, North Carolina State University, Raleigh, NC 27695, USA

**Abstract**

Externally generated wakes can impinge on porous layers in several practical heat transfer applications. However, the transport of these macroscale vortices and their resulting influence on heat transfer in the porous layer remain unknown. In this paper, two-dimensional pore-resolved simulations are used to examine a square-bluff-body wake impinging on a porous layer composed of an in-line array of heated square obstacles at $Re$ = 500–4000, $Pr = 0.7$ and $7.0$, and porosities $\phi$ = 0.75, 0.85, and 0.95. The incident macroscale vortices break down starting immediately at the porous-fluid interface. Following macroscale turbulent thermal transport in this entrance region, fluctuations farther downstream are aligned with the pore geometry and are generated at the microscale level by pore-throat shear, separation, and repeated obstacle wakes. Direct impact locally improves heat transfer at the porous-fluid interface by increasing the local Nusselt number relative to a non-impingement region. The interfacial heat-transfer enhancement varies with porosity, Reynolds number, and Prandtl number, with a peak enhancement of 18.2% observed for $\phi = 0.85$, $Pr = 7.0$, and $Re = 1000$. The wake-associated macroscale turbulent heat-flux contribution decays within a consistence entrance region extending approximately 3–4 unit cells into the porous layer across all investigated porosities, Reynolds numbers, and Prandtl numbers. For $Re \geq 1000$, the persistent wake momentum deficit subsequently causes the heat-transfer contrast to become negative, resulting in lower heat transfer in the impingement region than in the non-impingement region.

---

## 1 Introduction

The interaction between free-stream turbulence and porous layers is a fundamental problem in fluid mechanics with implications for industrial and environmental flows. A predictive understanding of vortex penetration, interfacial exchange of momentum and thermal energy, and the redistribution of unsteady flow structures at porous-fluid interfaces is particularly relevant to porous heat exchangers. These devices are often located downstream of flow obstructions and exposed to wakes generated by supports, ribs, or upstream components, where localized wake impingement can modify heat transfer at the entrance of the porous layer and downstream within the matrix. Similarly, free/porous-flow configurations arise in filtration systems [1,2], while flow through vegetative canopies is also relevant to wildfire transport and propagation [3]. In aerodynamic applications, porous and permeable surfaces have increasingly been explored for passive flow control, including wind-turbine surface treatments [4] and drag reduction via anisotropic permeability [5,6]. Bio-inspired surfaces provide another example of passive turbulent-flow control; seal-fur surfaces have been shown to reduce turbulent drag and streamwise turbulence intensity [7].

Beyond aerodynamic control, porous and microchannel architectures are also important in thermal-management systems, including microchannel heat sinks for electronics cooling [8]. In such systems, microscale vortices, vortex shedding, secondary flow instabilities, and pore-scale interactions with heated surfaces can substantially modify convective transport and the Nusselt-number response [9,10]. Porous lattice structures can also strongly enhance jet-impingement heat transfer [11]. At sufficiently large Reynolds numbers, inertial effects can produce substantial increases in pressure loss within regular porous structures [12]. While the impact of permeability on wall turbulence and porous-interface coupling has been extensively investigated in canonical channel and boundary-layer configurations [13,14], the local thermal consequence of direct wake impingement and its comparison with neighboring porous regions, that are not directly influenced by the wake remains insufficiently understood.

* Email address for correspondence: avkuznet@ncsu.edu

Despite the recognized influence of substrate permeability on total drag and turbulence modulation [15,16], relatively few studies have resolved the spatial evolution of wake-induced flow restructuring as macroscale vortices impinge on and are transported within a porous layer. Existing studies show that permeability strongly influences the intensity and spatial extent of the interaction between overlying turbulent flow and the porous medium and can promote Kelvin–Helmholtz-type interfacial instabilities [17]. However, the response of a porous layer to an externally imposed bluff-body wake remains insufficiently resolved. In particular, the spatial decay of wake-scale motion, its redistribution into pore-scale fluctuations, and the distinction between directly forced impingement regions and neighboring non-impingement baseline regions have not been systematically quantified. Understanding this gap is particularly important in thermal applications, where the hydrodynamic restructuring imposed by the porous matrix can affect heat transfer differently depending on the thermal diffusivity of the fluid.

The unresolved issue is whether an externally generated wake maintains its macroscale size and dynamics after encountering the first obstacle rows and whether any associated heat-transfer benefit persists downstream. A finite turbulent kinetic energy (TKE) inside the matrix cannot by itself demonstrate survival of the incident vortex because the pore geometry generates additional fluctuations through throat acceleration, shear-layer formation, separation, and vortex formation behind the solid obstacles. The persistence of the wake must therefore be assessed from the spatial organization and downstream evolution of the fluctuation field rather than from its magnitude alone.

This distinction motivates a two-stage thermal hypothesis. Direct wake impact can initially intensify local momentum and heat exchange around the first heated obstacles, producing an increased Nusselt number relative to a matched non-impingement baseline region. Farther downstream, the bluff-body wake-associated fluctuations weaken while the mean velocity deficit associated with the bluff-body persists within the impingement region. The same wake can therefore produce an entrance-region heat-transfer enhancement followed by a downstream deficit.

The present study tests this mechanism using two-dimensional pore-resolved simulations of a square-bluff-body wake impinging on a geometrically uniform in-line porous array. The impingement and non-impingement regions are evaluated using identical pore-scale control volumes and thermal boundary conditions, differing only in their exposure to the incoming wake. Porosity, Reynolds number, and Prandtl number are varied to determine how pore geometry, flow inertia, and thermal diffusivity govern the magnitude and downstream persistence of the wake-induced thermal response.

# 2 Solution Methods

## 2.1 Physical model and computational domain

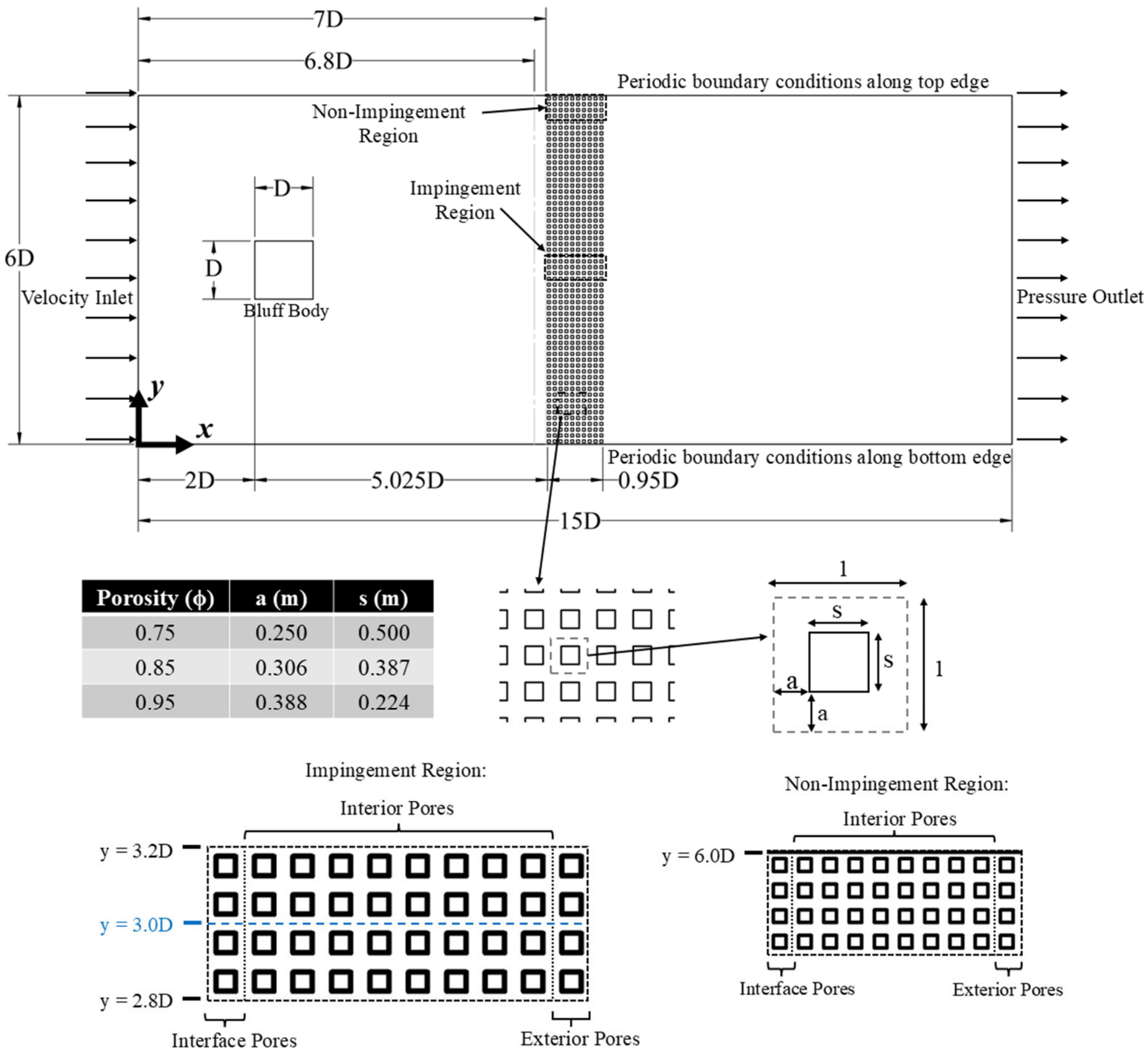

| Porosity (ϕ) | a (m) | s (m) |
|---|---|---|
| 0.75 | 0.250 | 0.500 |
| 0.85 | 0.306 | 0.387 |
| 0.95 | 0.388 | 0.224 |

Fig. 1: Computational domain and porous-layer sampling locations. A square bluff body of side length $D$ is placed upstream of an explicitly resolved in-line porous matrix. The full domain has dimensions $15D \times 6D$, and the porous layer begins at $X = x/D = 7.025$, corresponding to $5.025D$ downstream of the bluff-body leading edge. The lower insets show the pore geometry, the porosity-dependent obstacle size $s$, and the sampling locations used to distinguish the impingement and non-impingement regions. The impingement region corresponds to the portion of the porous matrix directly intercepted by the bluff-body wake, whereas the non-impingement region represents the baseline porous-layer response away from direct wake impact.

The computational domain contains a square bluff body of side length $D$ upstream of a homogeneous porous layer represented explicitly by a periodic in-line array of square obstacles. Resolving the individual obstacles allows pore-throat acceleration, separation, recirculation, and local wall heat transfer to be evaluated without a volume-averaged momentum closure.

The domain dimensions are $15D \times 6D$. A uniform inlet velocity $u_{in}$ is prescribed at the inlet and a gauge pressure of 0 Pa at the outlet. Periodic boundary conditions are applied at the upper and lower boundaries. The bluff body and porous obstacles satisfy no-slip conditions. The porous-obstacle surfaces are maintained at $T_w = 350\,K$, while the inlet fluid and bluff-body surface are at $T_{in} = 300\,K$. A backflow temperature of $300\,K$ is specified at the pressure outlet.

The porous matrix is described by the pore pitch $d$ (set equal to $1\,m$) and the square-obstacle side length $s$. The porosity is defined geometrically as the ratio of fluid volume to total volume within a representative pore cell. For an in-line array of square obstacles, this gives

$$\phi = 1 - \frac{s^2}{d^2} \tag{1}$$

The porosity values considered in this study are $\phi = 0.75$, 0.85, and 0.95. The corresponding obstacle size is obtained from

$$\frac{s}{d} = \sqrt{1 - \phi}, \tag{2}$$

and the open spacing around the obstacle is determined from the selected unit-cell geometry.

The flow conditions are characterized by a pore-scale Reynolds number based on the pore pitch *d* and the inlet velocity $u_{in}$:

$$Re = \frac{\rho u_{in} d}{\mu}. \tag{3}$$

The pore pitch *d* is used as the characteristic length scale for nondimensional governing equations, while $u_{in}$ is used as the reference velocity scale. The dimensionless spatial coordinates, velocity components, time, and pressure are therefore defined as

$$x^* = \frac{x}{d}, \quad y^* = \frac{y}{d}, \quad u^* = \frac{u}{u_{in}}, \quad v^* = \frac{v}{u_{in}}, \quad t^* = \frac{t u_{in}}{d}, \quad p^* = \frac{p}{\rho u_{in}^2}, \tag{4}$$

The governing equations are nondimensionalized with the pore pitch *d* and inlet velocity $u_{in}$, so x* = x/d and y* = y/d are used in the equations. For compact geometric presentation, figure axes use X = x/D and Y = y/D.

The temperature distribution is nondimensionalized using the imposed temperature difference between the heated porous obstacles and the incoming fluid:

$$\theta = \frac{T - T_{in}}{T_w - T_{in}} \tag{5}$$

With this definition, $\theta = 0$ at the inlet and $\theta = 1$ on the heated porous-obstacle surfaces. The inlet velocity $u_{in}$ is varied to achieve $Re = 500$, 1000, 2000, and 4000, allowing the influence of incoming wake strength and pore-scale inertia on thermal transport to be assessed systematically. Prandtl-number effects are investigated using $Pr = 0.7$ and $Pr = 7.0$.

Two matched sampling regions are used. The impingement region is centered on the bluff-body wake, and the non-impingement region is centered at the fixed lateral location shown in Fig. 1, outside the wake envelope. Each region contains the same number of pore-control volumes at identical streamwise stations and therefore has the same sampled fluid volume and heated perimeter within a given porosity. Differences between the regions are consequently attributed to wake exposure.

## 2.2 Numerical procedure and grid sensitivity

The simulations were performed in ANSYS Fluent 2025 R1 using a two-dimensional finite-volume formulation. The coupled algorithm was employed for pressure–velocity coupling, with the PRESTO! scheme used for pressure interpolation. Convection terms for momentum and energy were discretized using the QUICK scheme, while diffusion terms followed standard second-order central differencing based on least-squares cell gradients. A second-order implicit scheme was adopted for transient time integration.

Mesh sensitivity was assessed at $Re = 4000$, the condition with the strongest resolved unsteadiness among the simulated cases. The baseline mesh contained 4,133,760 elements. Refinement to 5,245,844 elements changed the time-averaged porous drag coefficient by 0.21% and the mean Nusselt number by 0.76%. Coarsening to 3,069,072 elements changed the same quantities by 0.92% and 1.06%, respectively. The baseline mesh was therefore retained for parametric study.

Table 1. Results of the Mesh Validation Study

| **Mesh Size Multiplier** | **Nodes** | **Global Size** | **Elements** | $\overline{C_D}$ | **Percent Error (%), Relative to 1.00 ×** | $\langle\overline{Nu}\rangle_V$ | **Percent Error (%), Relative to 1.00 ×** |
|---|---|---|---|---|---|---|---|
| 0.75 | 3784409 | 0.0825 | 5245844 | 11443.50 | 0.210 | 121.95 | 0.752 |
| 1.00 | 3225902 | 0.11 | 4133760 | 11467.64 | N/A | 121.04 | N/A |
| 1.25 | 2970672 | 0.1375 | 3626260 | 11452.43 | 0.133 | 121.59 | 0.454 |
| 1.50 | 2827735 | 0.165 | 3342356 | 11511.82 | 0.385 | 121.86 | 0.677 |
| 2.00 | 2689862 | 0.22 | 3069072 | 11572.99 | 0.919 | 122.32 | 1.058 |

## 2.3 Benchmark comparison with pore-resolved DNS data

After confirming mesh independence, an additional benchmark comparison was performed against the pore-resolved DNS data of Chu et al. (2019) [10], who investigated turbulent flow and heat transfer in a periodic array of square cylinders. The purpose of this comparison was to assess the extent to which the present two-dimensional pore-resolved framework reproduces the principal pore-scale mean-flow and thermal trends obtained from a fully three-dimensional DNS configuration. The quantities considered are the mean streamwise velocity, the streamwise velocity-fluctuation intensity, and the mean dimensionless temperature.

Figure 2 compares the present two-dimensional pore-resolved solution with the three-dimensional DNS data of Chu et al. [10]. The present model reproduces the principal spatial trends in the mean streamwise velocity, including the locations of obstacle-induced acceleration and deceleration, and captures the dominant distribution of the mean dimensionless temperature. Large discrepancies are observed in the streamwise velocity-fluctuation intensity, for which the two-dimensional calculations produce higher amplitudes at several locations. Such differences are expected because the present formulation does not represent spanwise flow structures, three-dimensional instabilities, or vortex stretching. Accordingly, this comparison is used as a benchmark of the mean pore-scale hydrodynamic and thermal behavior rather than as evidence of quantitative agreement with the complete three-dimensional turbulence statistics.

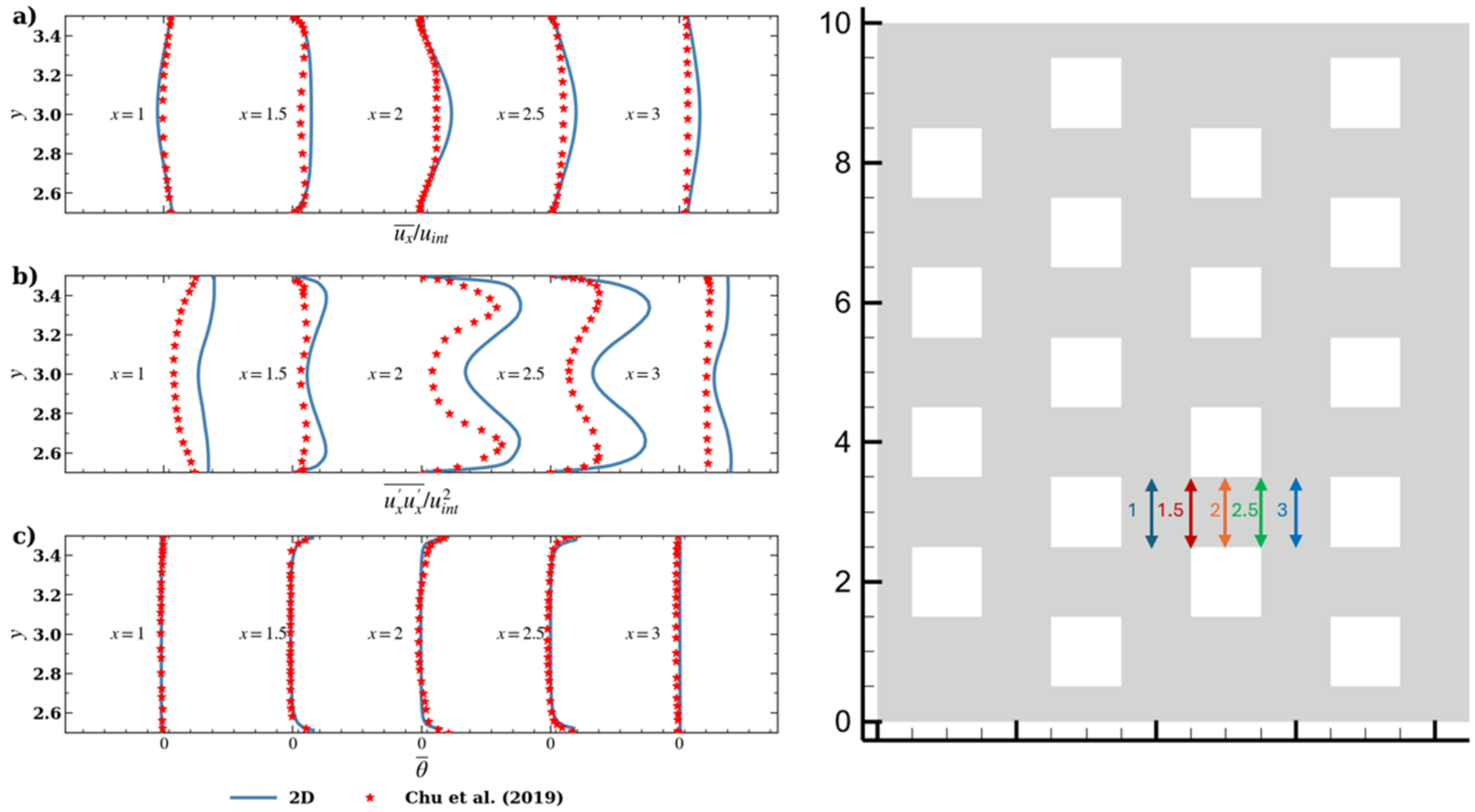

Fig. 2: Benchmark comparison of pore-scale statistics between the present two-dimensional simulations and the three-dimensional DNS data of Chu et al. (2019). The profiles are shown at several streamwise locations $x$ = 1, 1.5, 2, 2.5, and 3. The rows show: (a) mean streamwise velocity $\overline{u}_x/u_{int}$, (b) streamwise velocity-fluctuation intensity $\overline{u_x'u_x'}/u_{int}^2$, and (c) mean dimensionless temperature $\overline{\theta}$. The comparison shows that the present two-dimensional model captures the main pore-scale trends in mean flow and temperature, while magnitude differences reflect the absence of three-dimensional turbulence mechanisms in the two-dimensional formulation.

### 2.4 Governing equations and thermal transport metrics

Fluid motion and heat transfer in the fluid region are governed by the incompressible Navier–Stokes equations and the energy equation, which together describe the coupled evolution of the velocity, pressure, and temperature fields:

$$\nabla^* \cdot \mathbf{u}^* = 0, \tag{6}$$

$$\frac{\partial \mathbf{u}^*}{\partial t^*} + (\mathbf{u}^* \cdot \nabla^*)\mathbf{u}^* = -\nabla^* p^* + \frac{1}{Re}\nabla^{*2}\mathbf{u}^*, \tag{7}$$

$$\frac{\partial \theta}{\partial t^*} + (\mathbf{u}^* \cdot \nabla^*)\theta = \frac{1}{Re \cdot Pr}\nabla^{*2}\theta, \tag{8}$$

where $\mathbf{u}^*$ is the dimensionless velocity vector, $p^*$ is the dimensionless pressure, $\theta$ is the dimensionless temperature, and $\nabla^*$ is the gradient operator with respect to the dimensionless spatial coordinates $(x^*, y^*)$. The Prandtl number is defined as

$$Pr = \frac{\nu}{\alpha} \tag{9}$$

where the thermal diffusivity α was selected to obtain *Pr = 0.7* and *Pr = 7.0*, representative of commonly used working fluids such as air and water, respectively.

To characterize wake-induced unsteadiness in the two-dimensional field, the dimensionless turbulent kinetic energy is defined as

$$\langle TKE \rangle_V = \frac{1}{2}\frac{\langle \overline{u'^2} \rangle_V + \langle \overline{v'^2} \rangle_V}{u_{in}^2}, \tag{10}$$

where $u'$ and $v'$ are the velocity fluctuations about the Reynolds-averaged streamwise and transverse velocity components, respectively. This quantity is used to identify regions where velocity-fluctuation intensity is concentrated or redistributed within the porous layer.

To quantify the heat-transfer response of the porous structure, the local wall heat flux is obtained from the temperature gradient at the solid boundary. In the present work, the reported mean Nusselt number is defined using the local fluid temperature in the sampled pore volume as

$$\langle \overline{Nu} \rangle_V = \frac{\langle \overline{q}_w'' \rangle_S \cdot s}{k\left(T_w - \langle \overline{T}_f \rangle_V\right)} = \frac{\langle \frac{\partial \theta}{\partial n^*}\Big|_w \rangle_S \cdot \left(\frac{s}{d}\right)}{1 - \langle \overline{\theta}_f \rangle_V}, \tag{11}$$

where $n^* = n/d$ is the dimensionless coordinate normal to the solid boundary, and $\langle \overline{\theta}_f \rangle_V$ is the intrinsic volume-averaged mean dimensionless fluid temperature.

For comparison, an inlet-temperature-based Nusselt number can be defined as

$$Nu_{in} = \frac{\langle \overline{q}_w'' \rangle_S \cdot s}{k\,(T_w - T_{in})} \tag{12}$$

Unlike the local-temperature-based mean Nusselt number, $Nu_{in}$ uses the fixed reference temperature difference $T_w - T_{in}$ as the reference scale. It is therefore directly proportional to the dimensional wall heat flux and provides a complementary measure of the absolute heat-transfer rate.

To investigate the heat-transfer mechanisms inside the porous region from a macroscopic perspective, the local energy equation is expressed in double-averaged (Reynolds- and volume-averaged) form. The analysis is restricted to a two-dimensional configuration, retaining only the streamwise and transverse directions. After averaging, the fluid-phase double-averaged energy equation is written as

$$\phi\langle\bar{u}_j^*\rangle\frac{\partial\langle\bar{\theta}_f\rangle_V}{\partial x_j^*} = \frac{1}{Pr\,\mathrm{Re}}\frac{\partial}{\partial x_j^*}\left(\frac{\partial\,\phi\langle\bar{\theta}_f\rangle_V}{\partial x_j^*} + \frac{1}{\Delta V^*}\int_A n_j\,\bar{\theta}_f\,dA^*\right) - \frac{\partial}{\partial x_j^*}\phi\left\langle\overline{u_j^{*\prime}\theta_f^{\prime}}\right\rangle_V - \frac{\partial}{\partial x_j^*}\phi\langle\tilde{u}_j^*\tilde{\theta}_f\rangle_V + \frac{1}{Pr\,Re\,\Delta V^*}\int_{A^*} n_j\frac{\partial\bar{\theta}_f}{\partial x_j^*}\,dA^*, \qquad j = 1,2, \tag{13}$$

where $\phi$ is the porosity, $\Delta V^*$ is the averaging control volume, and $A^*$ is the fluid–solid interfacial boundary contained within that control volume. The overbar denotes Reynolds averaging, $\langle\cdot\rangle_V$ denotes volume averaging, $\langle\cdot\rangle_S$ denotes surface averaging, and the tilde denotes the spatial deviation from the intrinsic volume-averaged quantity. In the present two-dimensional implementation, the averaging volume is interpreted as a representative control volume of unit out-of-plane depth, so that

$$\Delta V^* = \Delta x^* \cdot \Delta y^*. \tag{14}$$

After Reynolds and volume averaging, Eq. (13) yields the thermal-budget decomposition used in the present analysis. For the two-dimensional formulation, this balance is written as

$$C_p^T = D_p^{\nu T} + H_p^t + H_p^d + S_w\,, \tag{15}$$

where the individual terms are defined as follows. The mean convective transport term,

$$C_p^T = \phi\left(\langle\bar{u}^*\rangle\frac{\partial\langle\bar{\theta}_f\rangle_V}{\partial x^*} + \langle\bar{v}^*\rangle\frac{\partial\langle\bar{\theta}_f\rangle_V}{\partial y^*}\right) \tag{16}$$

represents transport of the mean temperature field by the intrinsic volume-averaged mean velocity. The diffusive transport term,

$$D_p^{\nu T} = \frac{1}{Pr\,Re}\left[\frac{\partial^2}{\partial x^{*2}}\left(\phi\langle\bar{\theta}_f\rangle_V\right) + \frac{\partial^2}{\partial y^{*2}}\left(\phi\langle\bar{\theta}_f\rangle_V\right)\right] + \frac{1}{Pr\,Re}\left(\frac{\partial I_x}{\partial x^*} + \frac{\partial I_y}{\partial y^*}\right), \tag{17}$$

contains both the molecular diffusion contribution and an interfacial tortuosity contribution arising from the fluid–solid interface geometry within the averaging volume. The tortuosity terms are defined as

$$I_x = \frac{1}{\Delta V^*}\int_{A^*} n_x\,\bar{\theta}_f\,dA^*, \qquad I_y = \frac{1}{\Delta V^*}\int_{A^*} n_y\,\bar{\theta}_f\,dA^*. \tag{18}$$

The turbulent heat-flux term,

$$H_p^t = -\left[\frac{\partial}{\partial x^*}\left(\phi\left\langle\overline{u^{*\prime}\theta_f^{\prime}}\right\rangle_V\right) + \frac{\partial}{\partial y^*}\left(\phi\left\langle\overline{v^{*\prime}\theta_f^{\prime}}\right\rangle_V\right)\right], \tag{19}$$

accounts for heat transport associated with fluctuations of velocity and temperature about their Reynolds-averaged values. The dispersive heat-flux term,

$$H_p^d = -\left[\frac{\partial}{\partial x^*}\left(\phi\langle\tilde{u}^*\,\tilde{\theta}_f\rangle_V\right) + \frac{\partial}{\partial y^*}\left(\phi\langle\tilde{v}^*\,\tilde{\theta}_f\rangle_V\right)\right], \tag{20}$$

represents heat transport generated by spatial deviations of the local velocity and temperature fields from their intrinsic volume-averaged values within the pore space. The wall heat-transfer term,

$$S_w = \frac{1}{Pr\,Re\,\Delta V^*}\int_A\left(n_x\frac{\partial\bar{\theta}_f}{\partial x^*} + n_y\frac{\partial\bar{\theta}_f}{\partial y^*}\right)dA^* \tag{21}$$

captures the direct contribution of wall heat-transfer associated with the conductive temperature gradient at the fluid–solid boundary. This decomposition is used throughout the analysis to identify how wake-induced interfacial fluctuations and pore-scale flow restructuring influence thermal transport within the porous layer. Finally, the residual is defined as

$$\mathrm{R} = D_p^{\nu T} + H_p^t + H_p^d + S_w - C_p^T \tag{22}$$

## 2.5 Impingement and non-impingement comparison metrics

To quantify the thermal consequence of the direct wake impingement independently of the overall heat-transfer level, the local-temperature-based Nusselt number defined above is compared between the impingement and non-impingement sampling regions. Let $Nu_r(X_j)$ denote the Nusselt number averaged over the sampled pore control volumes belonging to region *r* at streamwise position $X_j$, where *r* = imp and *r* = non represent the impingement and non-impingement regions, respectively. It is calculated as

$$\mathrm{Nu_r}(X_j) = \frac{\left[\sum_{m\in\Omega_{r,j}} \langle \overline{Nu} \rangle_{V,m}\, \Delta V^f_{,m}\right]}{\sum_{m\in\Omega_{r,j}} \Delta V^f_{,m}}\ , r \in \{imp, non\}, \tag{23}$$

where $\Omega_{r,j}$ is the set of sampling pore-control volumes in region *r* and streamwise window *j*, and $\Delta V^f_{,m}$ is the corresponding fluid volume. Because the pore-control volumes have equal dimensions within each porosity case, Eq. (23) reduces to:

$$\mathrm{Nu_r}(X_j) = \frac{1}{N_{r,j}} \sum_{m\in\Omega_{r,j}} \langle \overline{Nu} \rangle_{V,m}\ , \tag{24}$$

$$\Delta\mathrm{Nu}(\mathrm{X_j}) = \mathrm{Nu_{imp}}(\mathrm{X_j}) - \mathrm{Nu_{non}}(\mathrm{X_j}) \tag{25}$$

Positive $\Delta Nu$ indicates heat transfer enhancement; negative $\Delta Nu$ indicates diminished heat transfer.

$$\mathrm{E}(X_\mathrm{j}) = \frac{\Delta\mathrm{Nu}(\mathrm{X_j})}{\mathrm{Nu_{non}}(\mathrm{X_j})} \times 100\%\ , \mathrm{E_{int}} = \mathrm{E}(\mathrm{X_{int}}) \tag{26}$$

where $\mathrm{X_{int}}$ denotes the porous-interface sampling window, *E(X)* is the local relative impingement contrast along the porous layer, and $E_{int} = E(X_{int})$ is its interface value.

When $\Delta Nu(X_j) > 0$ and $\Delta Nu(X_{j+1}) \leq 0$, estimate the crossover by linear interpolation:

$$\mathrm{X_c} = \mathrm{X_j} + \left[\frac{\Delta\mathrm{Nu}(\mathrm{X_j})}{\Delta\mathrm{Nu}(\mathrm{X_j}) - \Delta\mathrm{Nu}(\mathrm{X_{j+1}})}\right](\mathrm{X_{j+1}} - \mathrm{X_j}) \tag{27}$$

To quantify the mean-flow signature remaining along the wake-affected region, the time-averaged streamwise velocity is averaged over the same matched fluid pore-control volumes used for the Nusselt-number comparison. The local relative mean-velocity deficit is defined as

$$\mathrm{D_U}(X_j) = \frac{\overline{\mathrm{U}}_{non}(X_j) - \overline{\mathrm{U}}_{\mathrm{imp}}(X_j)}{\overline{\mathrm{U}}_{non}(X_j)} \times 100\%\ , \tag{28}$$

Positive $D_U$ indicates that the impingement region has a lower mean streamwise velocity than the matched non-impingement region. A value of $D_U = 0$ indicates no mean-velocity difference between the two regions, whereas $D_U < 0$ indicates a higher mean streamwise velocity in the impingement region. Because $D_U$ and $\Delta Nu$ are evaluated over identical pore-control volumes and streamwise windows, their downstream evolution can be compared directly. Two additional no-bluff-body control simulations were performed at *Re = 1000*, *Pr = 7.0*, and $\phi = 0.75$ using laminar and synthetic-vortex inlet conditions. These controls use the same porous geometry, thermal boundary conditions, and pore-control-volume averaging procedure as the reference case and are used to distinguish externally supplied inlet-fluctuation effects from fluctuations generated intrinsically by the porous matrix. The control configuration, comparison metrics, and results are provided in the Supplementary Material.

## 3 Results and Discussion

The impingement of externally generated turbulence at the porous-fluid interface is ubiquitous in practical applications of porous media. When turbulent vortices are generated outside of the porous layer, their length scale often exceeds the pore scale, resulting in a macroscale-to-microscale interaction at the porous-fluid interface. As a result, turbulence transport at the porous-fluid interface is heterogeneous and strongly influenced by the impinging bluff-body wake. The present study of this mechanism is organized around the hypothesis of a two-stage flow behavior: macroscale turbulent transport at the porous-fluid interface followed by predominantly microscale transport deep inside the porous layer.

### 3.1 Thermal response in the impingement and non-impingement regions

When a macroscale wake structure impinges on a porous layer, its characteristic length scale is larger than the pore spacing. The incoming vortical motion therefore encounters a sudden geometric restriction at the porous-fluid interface, where repeated obstacle interactions redistribute the imposed large-scale motion toward smaller pore-scale motions. Our previous work showed that this hydrodynamic restructuring is accompanied by elevated volume-averaged TKE and increased TKE dissipation near the porous-fluid interface [18]. This hydrodynamic restructuring raises the central heat-transfer question addressed here: does the additional fluctuation energy introduced by direct wake impingement produce a corresponding enhancement of wall heat flux and Nusselt number, and how far does that thermal influence persist inside the porous layer?

The time-averaged streamwise velocity field in Fig. 3 shows that the bluff-body wake reaches the porous-fluid interface and enters the porous layer along the impingement region. The corresponding interfacial thermal response is first established before interpreting the downstream fluctuation field for the reference case, $Re = 1000$, $Pr = 7.0$, and $\phi = 0.75$. Direct impingement modifies the thermal field at the first heated obstacle rows before the flow is reorganized by the entrance region of the porous matrix. Figure 4(a) shows a heterogeneous temperature distribution within the porous matrix, with the bulk fluid temperature in the impingement region transitioning from below to above that in the non-impingement region. The stronger wall-normal temperature gradient produced at the impingement location increases the surface-averaged wall heat flux from ~3000 W m$^{-2}$ in the non-impingement region to ~3480 W m$^{-2}$ in the impingement region [Fig. 4(b)], an increase of about 16%. The corresponding volume-averaged Nusselt number, calculated using Eq. (23), increases from 33.2 in the non-impingement to 37.4 in the impingement region at the interface, corresponding to an interfacial heat transfer enhancement $E_{int} = 12.7\%$ [Fig. 4(c)]. The separate calculations of the heat-flux and Nusselt-number enhancements provide a useful distinction in heat-transfer mechanism. The wall heat flux measures the absolute thermal transfer, whereas the local-temperature-based Nusselt number accounts for the increase in local volume-averaged fluid temperature as the fluid is heated while moving deeper into the porous layer. The Nusselt number is therefore used as the principal wake-specific metric because it isolates the convective response from the changing local thermal driving difference, while wall heat flux represents the absolute thermal transfer.

The heat transfer enhancement observed with the Nusselt number distributions closely follows the TKE inside the porous matrix, where the peak volume-averaged TKE is ~0.42 in the impingement region compared with ~0.11 in the matched non-impingement region [Fig. 4(d)]. TKE is initially introduced into the porous layer at the impingement region by the bluff-body wake. Within the porous matrix, the obstacle geometry produces TKE at the microscale through pore-throat acceleration, shear-layer formation, separation, and repeated obstacle wakes. This is evident in Fig. 5, which shows that TKE is elevated deep inside the porous layer in both the impingement and non-impingement regions. The accompanying temperature field shows progressive fluid heating downstream, while the wall-heat-flux and Nusselt-number fields show the corresponding spatial heat-transfer response downstream. Thus, these results support the notion of a two-stage mechanism of TKE transport, where initially macroscale turbulence is succeeded by production at the microscale level. Our previous work demonstrated with two-point velocity fluctuation autocorrelation functions that while the vortical structures at the impingement region of the porous-fluid interface are larger than the pore scale, the structures deep inside the porous layer are of microscale size [18].

A strong local association between fluctuation intensity and heat transfer emerges in the reference case (Fig. 6), for which the linear fit gives $R^2 = 0.910$. This correlation demonstrates that locations with stronger velocity fluctuations generally exhibit stronger convective heat transfer in the reference case. It does not, however, establish that the bluff-body wake survives inside the matrix. That distinction becomes clear downstream: the volume-averaged TKE increases to approximately 0.65–0.70 in the interior pore blocks, while the enhancement of heat transfer by impingement measured by $\Delta Nu$ decreases to zero and then becomes negative. The finite TKE observed farther downstream therefore reflects obstacle-generated pore-scale production that occurs increasingly homogeneously in both impingement and non-impingement regions, whereas $\Delta Nu$ isolates the additional thermal effect of direct wake exposure.

Additional no-bluff-body inlet-condition controls were used to test whether substantial downstream fluctuation energy necessarily reflects persistence of externally supplied fluctuations (details are provided in the Supplementary Material). Considering the case of $Re = 1000$, $Pr = 7.0$, and $\phi = 0.75$, a synthetic-vortex inlet was configured to provide a TKE magnitude at the porous entrance comparable to that measured in the bluff-body impingement region, whereas the laminar-inlet control entered the matrix as laminar flow. Despite this difference, substantial TKE developed downstream in the laminar-inlet case, while the differences in TKE, Nusselt number, and surface-averaged wall heat flux between the two controls progressively diminished. These controls independently support the interpretation that finite downstream TKE can be generated intrinsically by the porous geometry, which homogenizes TKE transport and diminishes any wake-induced heat transfer enhancement.

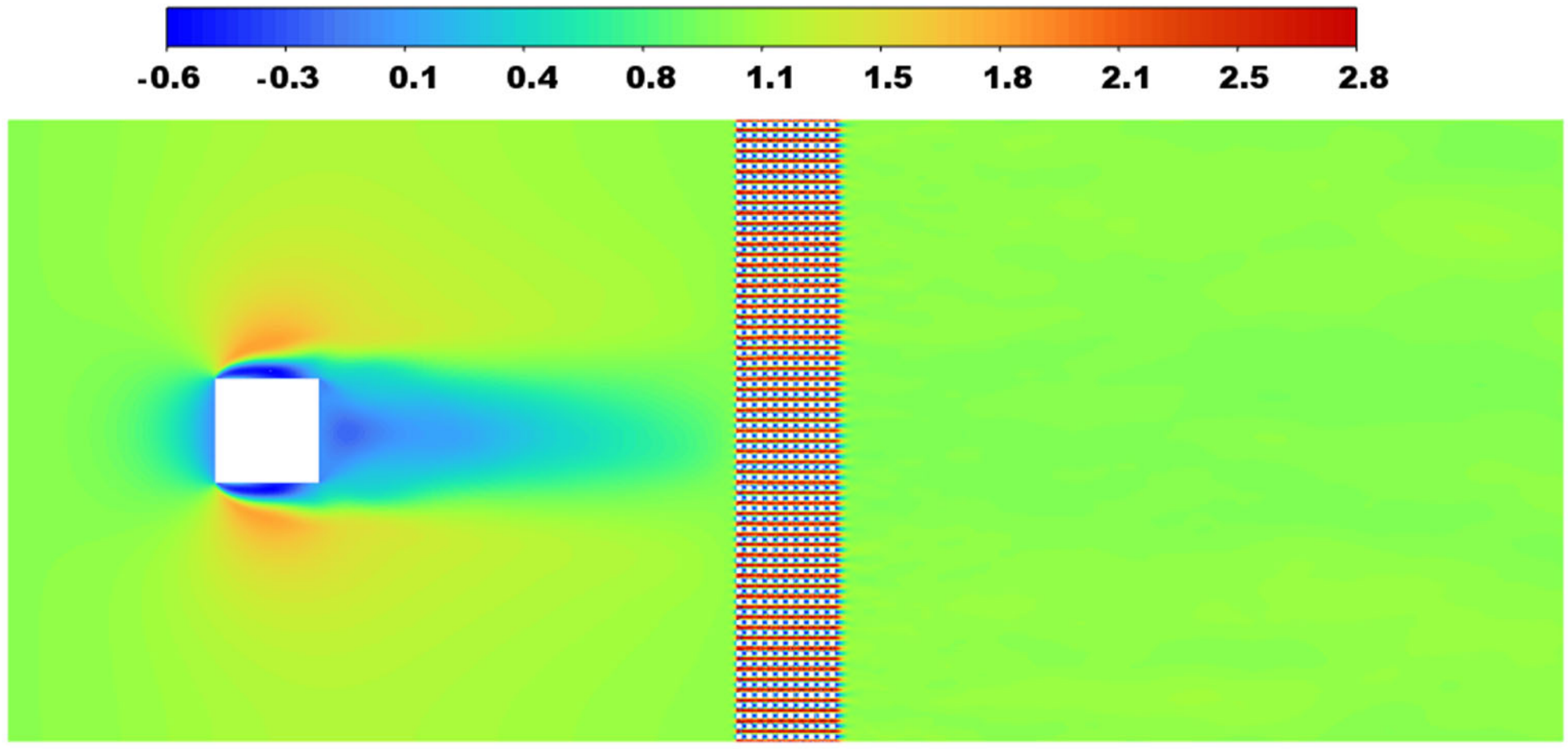

Fig. 3: Time-averaged streamwise velocity field for the square-bluff-body wake and downstream porous layer at $Re = 1000$, $Pr = 7.0$, and $\phi = 0.75$. The impingement region is the central sampling band directly intersected by the low-velocity bluff-body wake, whereas the non-impingement region is the geometrically matched lateral band located outside the main wake footprint, as defined in Fig. 1. The wake reaches the porous entrance and remains centered on the impingement region. Negative streamwise velocities immediately downstream of the bluff body identify the recirculation zone.

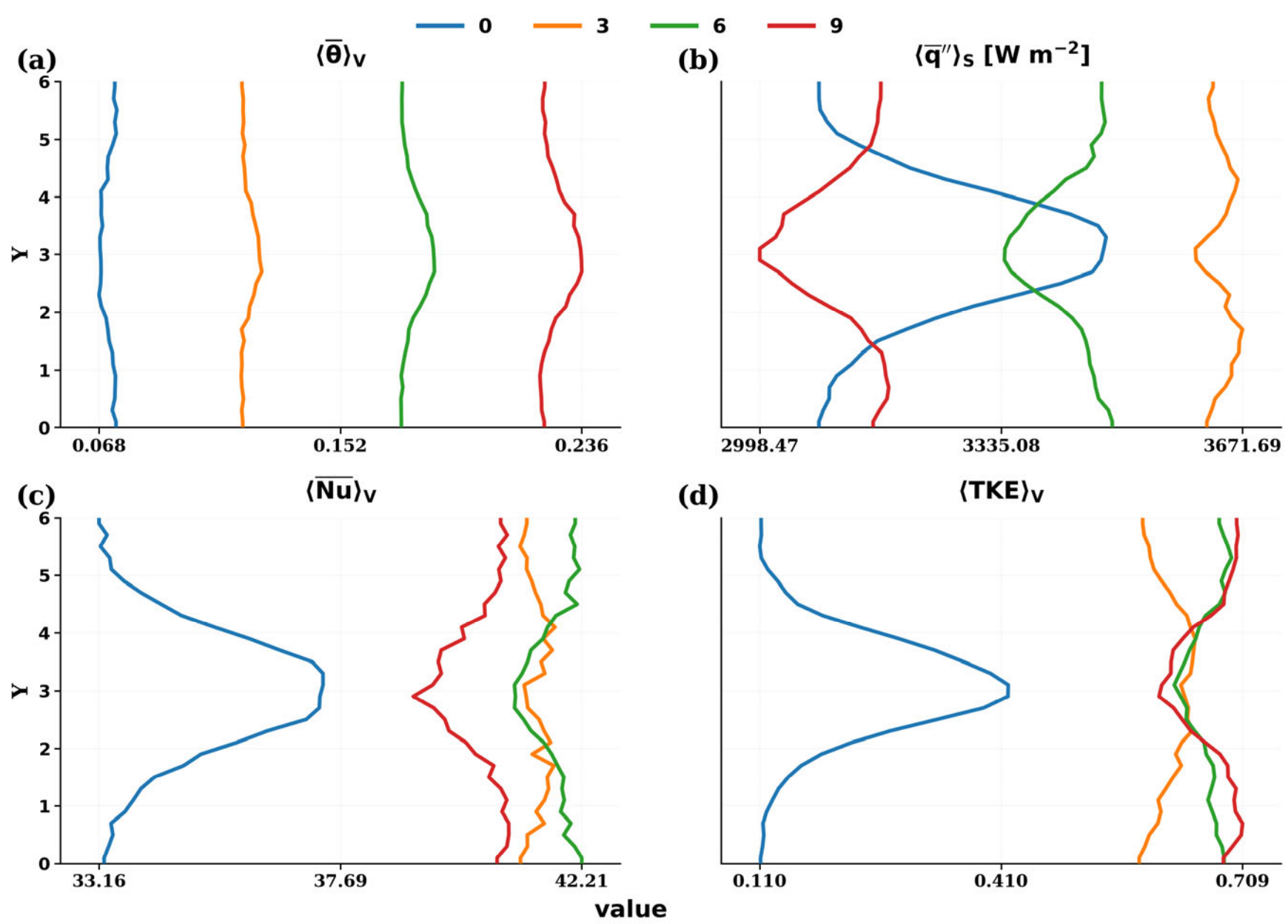

Fig. 4: Vertical profiles of (a) volume-averaged mean dimensionless fluid temperature, (b) surface-averaged wall heat flux, (c) local-temperature-based mean Nusselt number, and (d) volume-averaged turbulent kinetic energy at pore-column locations 0, 3, 6, and 9 for $Re = 1000$, $Pr = 7.0$, and $\phi = 0.75$. At the interface, the impingement centerline exhibits higher wall heat flux, Nusselt number, and TKE than the matched non-impingement region. Downstream, TKE remains finite while the wake-specific Nusselt-number advantage disappears.

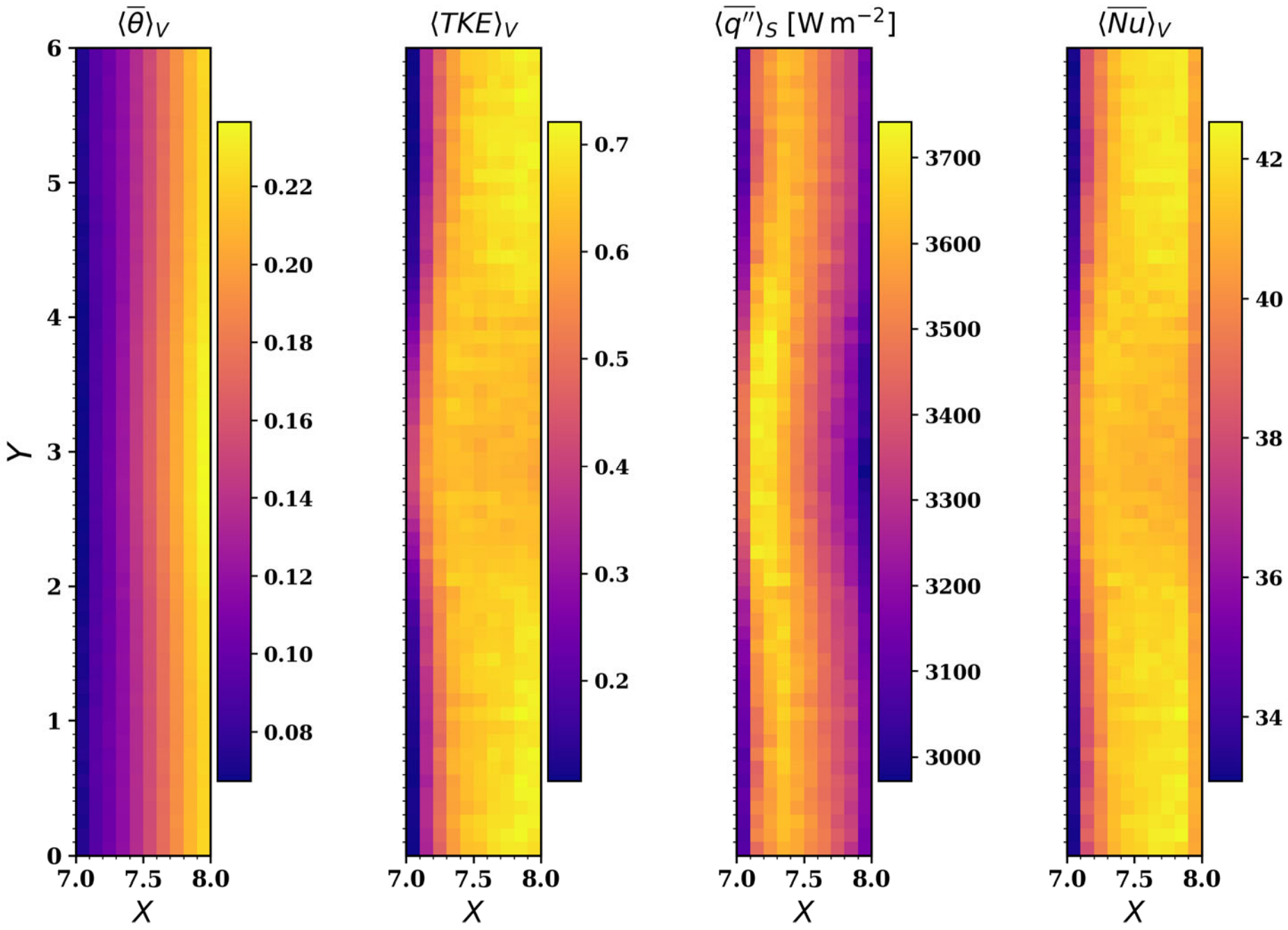

Fig. 5: Spatial distributions of volume-averaged dimensionless temperature, turbulent kinetic energy, surface-averaged wall heat flux, and local-temperature-based mean Nusselt number within the porous layer, $7.0 \leq X \leq 8.0$, for $Re = 1000$, $Pr = 7.0$, and $\phi = 0.75$. The TKE field becomes obstacle-locked inside the matrix, whereas the largest wall heat flux remains concentrated near the entrance.

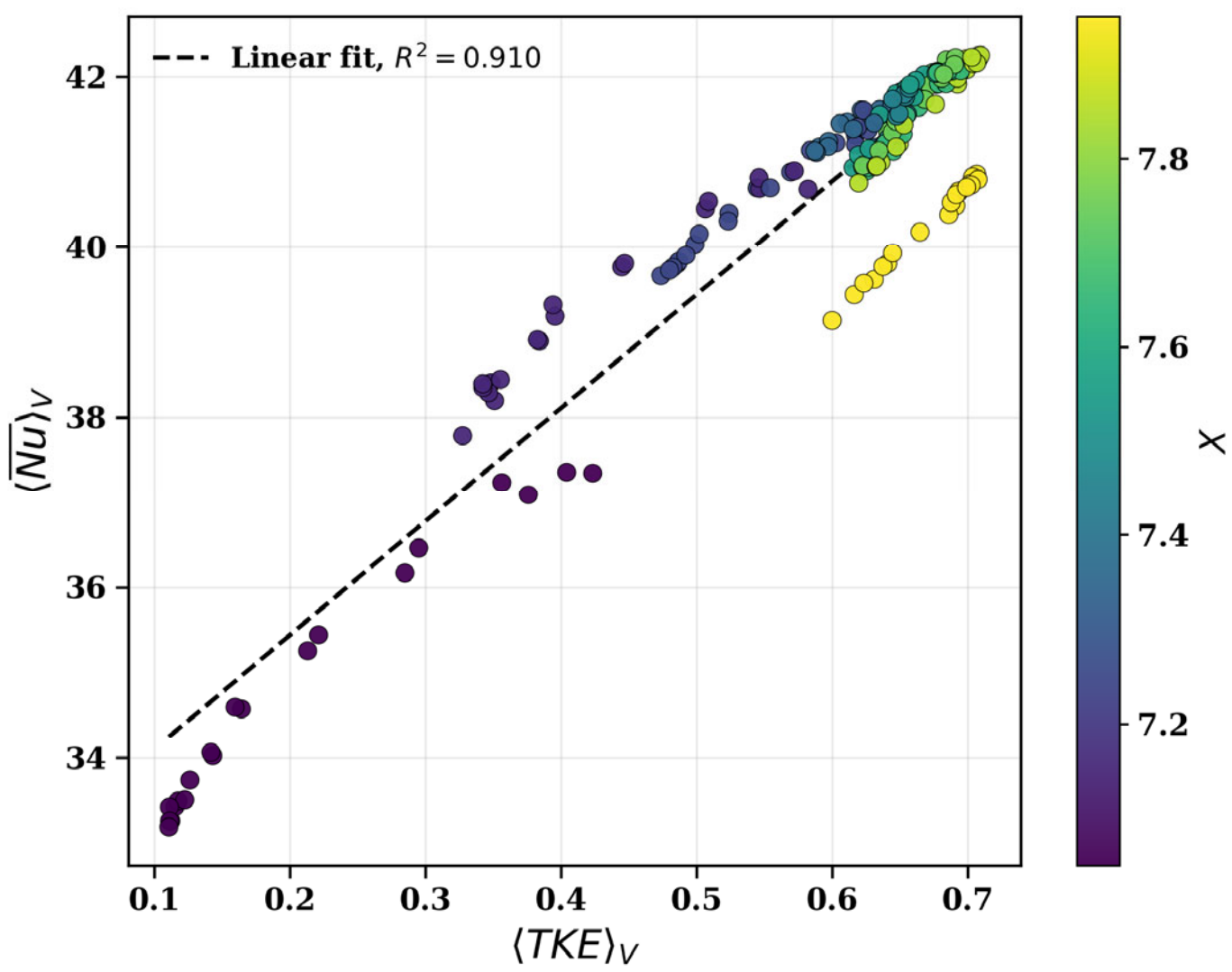

Fig. 6: Relation between volume-averaged turbulent kinetic energy and local-temperature-based mean Nusselt number for the reference case $Re = 1000$, $Pr = 7.0$, and $\phi = 0.75$. The linear fit gives $R^2 = 0.910$, showing a strong local association between fluctuation intensity and convective exchange. The correlation is not used as evidence that the incident wake-scale vortex survives inside the porous matrix.

The thermal energy transport decomposition explains why the heat transfer enhancement in the impingement region is localized (Fig. 7). The wall heat flux source term $S_w$ represents thermal energy entering the fluid from the heated solid surface. The mean-convective term $C_p^T$ is the advective response to that wall input $S_w$, while $H_p^t$ and $H_p^d$ describe redistribution by fluctuations (turbulent) and pore-scale spatial deviations (dispersion), respectively. Near the porous-fluid interface, the impingement region exhibits greater macroscale thermal transport through $H_p^t$ and $H_p^d$ relative to the non-impingement region, showing that direct wake exposure modifies how the thermal energy introduced at the solid obstacle walls is transported away from the first obstacle rows. Note that the volume-averaged diffusion term is much smaller than the most significant terms because it represents transport by macroscale temperature gradients.

Farther downstream, the turbulent heat-flux contribution $H_p^t$ decreases to approximately zero, indicating that the influence of the impinging macroscale wake has diminished. In contrast, the dispersive heat flux remains nonzero and becomes uniform in the streamwise direction because the fluid continues to be heated by successive solid obstacles, which sustains pore-scale spatial deviations of temperature along the flow direction. The additional thermal redistribution associated specifically with the incident macroscale wake is therefore completed, while pore-scale dispersive transport and wall heat transfer continue downstream. The corresponding evolution of the mean-velocity deficit and Nusselt-number contrast is quantified separately in Section 3.2.2 using Fig. 10.

Taken together, these results support the proposed two-stage thermal transport hypothesis. Direct vortex impingement initially increases wall-normal momentum and thermal transport near the first heated obstacle rows, thereby increasing the wall-normal temperature gradient, wall heat flux, and Nusselt number at the porous-fluid interface. Farther downstream, the wake-specific turbulent contribution vanishes, and the additional dispersive contribution relaxes toward the pore-generated background. Meanwhile, the mean-velocity deficit persists, ultimately producing a heat-transfer deficit relative to the non-impingement region.

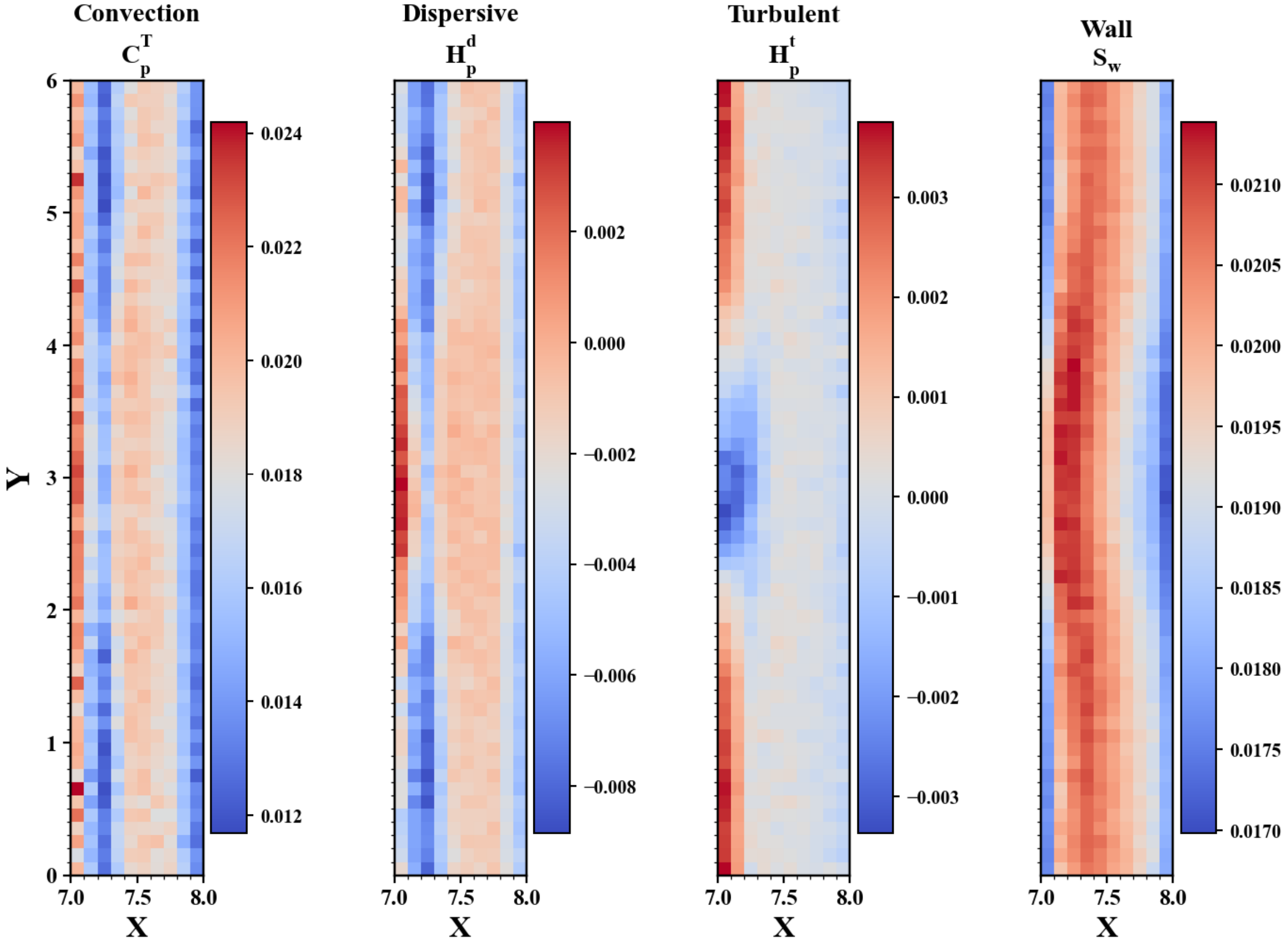

Fig. 7: Spatial distributions of the dimensionless thermal-budget terms for $Re = 1000$, $Pr = 7.0$, and $\phi = 0.75$: mean convection $C_p^T$, dispersive redistribution $H_p^d$, turbulent redistribution $H_p^t$, and wall input $S_W$. The wall-input term represents thermal energy transferred from the heated solid to the fluid, while the mean-convection term represents the corresponding mean-advective response. Differences between the impingement and non-impingement regions are concentrated near the porous-layer entrance and diminish downstream. The volume-averaged diffusion term is omitted because its contribution is small relative to the other budget terms.

## 3.2 Parametric Study on the Impingement/Non-Impingement Thermal Response

After establishing the mechanism of thermal energy transport during macroscale vortex impingement at a porous-fluid interface, the findings are re-evaluated to quantify how pore geometry, flow inertia, and thermal diffusivity modify the contrasting heat transfer behavior in the impingement versus non-impingement regions. The discussion is organized around porosity, Reynolds number, and Prandtl number because these parameters control the relative roles of pore-scale advection, wall heat exchange, and turbulent/dispersive redistribution in the present simulations.

### 3.2.1 Effect of Porosity

Porosity is varied to determine how obstacle blockage and pore opening modify the interaction between the incident wake and the heated matrix. At $\phi = 0.75$, the larger obstacles and narrower passages produce stronger confinement and stronger obstacle-generated shear in both sampling regions, which reduces the relative difference associated specifically with direct wake exposure. At $\phi = 0.95$, the smaller obstacles and wider passages weaken the interaction between the incoming wake and the heated solid surfaces. The intermediate porosity, $\phi = 0.85$, provides a balance between wake exposure and obstacle-mediated wall exchange, consistent with the largest relative interface heat-transfer enhancement shown in Fig. 9.

The comparison in Fig. 8 also shows that the difference between the impingement and non-impingement fields becomes less pronounced farther downstream for all three porosities. This indicates that the direct influence of the incoming wake is strongest near the porous-fluid interface, while the downstream response becomes increasingly similar between the two matched regions. The quantitative consequence of this evolution is examined in Fig. 9 through $\Delta Nu(X)$ and $E(X)$.

Figure 9 reveals a non-monotonic dependence of the wake-induced interface enhancement on porosity. The largest relative enhancement occurs at the intermediate porosity, $\phi = 0.85$, where $E_{int} = 18.2\%$, compared with $12.7\%$ at $\phi = 0.75$ and $15.5\%$ at $\phi = 0.95$. This result shows that increasing pore opening does not monotonically increase the thermal benefit of direct impingement. The strongest wake-specific response occurs at the intermediate geometry, where the difference between the impingement and non-impingement regions is greatest.

The porosity comparison also separates the magnitude of the heat transfer enhancement at the porous-fluid interface from the crossover location $X_c$, where $\Delta Nu$ changes from positive to negative. Although $\phi = 0.85$ produces the largest interface enhancement, the crossover location remains essentially unchanged among the three porosities. For each case, $X_c$ is obtained from the linear interpolation defined in Eq. (27) between the last station with $\Delta Nu > 0$ and the first station with $\Delta Nu \leq 0$. The resulting crossover locations cluster around $Xc \approx 7.35$ for $\phi$ = 0.75, 0.85, and 0.95. Within the present streamwise sampling resolution, no systematic shift of $X_c$ with porosity can be resolved.

At the exit of the porous layer, the impingement-region Nusselt number is lower than the non-impingement value by approximately 3.4%, 0.9%, and 5.5% for $\phi$ = 0.75, 0.85, and 0.95, respectively. Among the tested geometries, $\phi$ = 0.85 therefore provides both the largest interface enhancement and the smallest downstream relative heat-transfer deficit. These results indicate that, over the porosity range examined here, porosity primarily modifies the magnitudes of both the heat transfer enhancement at the porous-fluid interface and the downstream deficit within the porous layer.

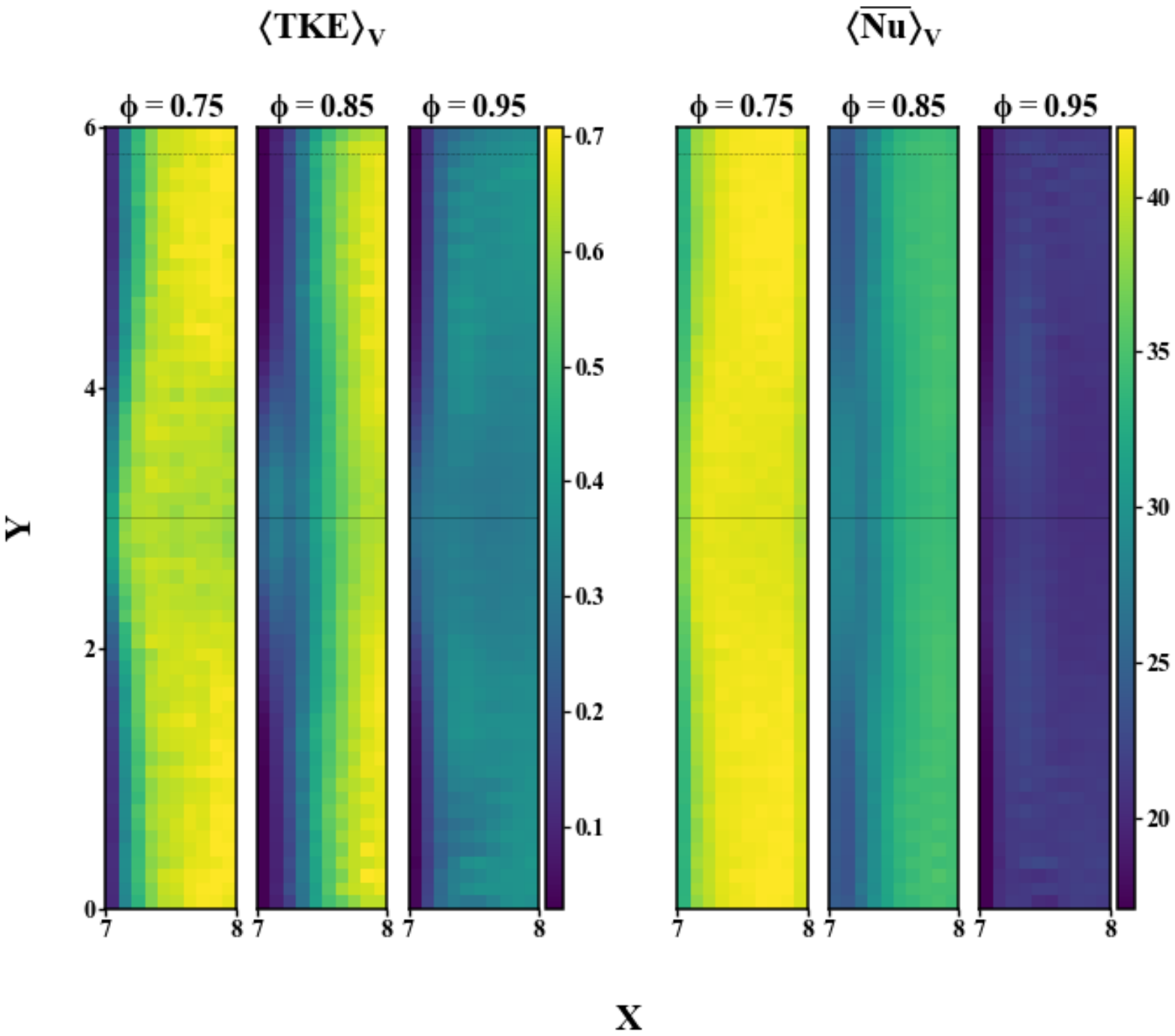

Fig. 8: Porosity dependence of volume-averaged TKE and local-temperature-based Nusselt-number fields at $Re = 1000$ and $Pr = 7.0$ for $\phi = 0.75$, 0.85, and 0.95. Horizontal guides identify the matched impingement and non-impingement sampling regions. Porosity modifies both the fluctuation intensity and the corresponding thermal response.

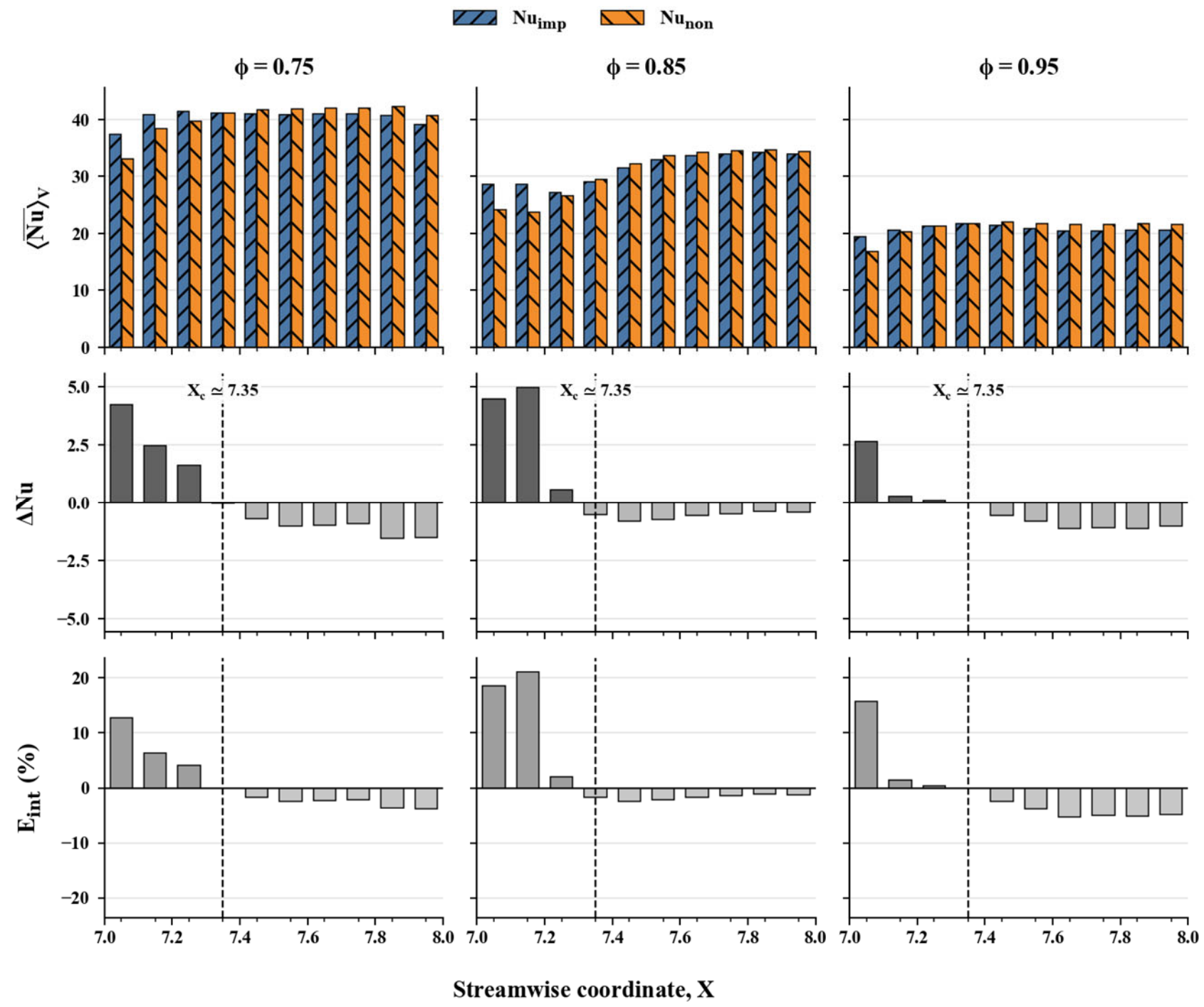

Fig. 9: Streamwise evolution of the region-averaged local-temperature-based Nusselt number for the matched impingement and non-impingement regions at $\phi = 0.75$, 0.85, and 0.95, $Re = 1000$, and $Pr = 7.0$. At each $X$ station, $Nu_r$ is the fluid-volume-weighted average of the pore-control-volume Nusselt numbers in the corresponding sampling region, as defined in Eq. (23); equal pore-control volumes reduce this to an arithmetic average. The second and third rows show $\Delta Nu(X)$ and $E(X)$, respectively. The interpolated zero crossing is $X_c \approx 7.35$ for all three porosities.

### 3.2.2 Effect of Reynolds and Prandtl Numbers on the Impingement/Non-Impingement Thermal Contrast

Reynolds number controls the relative importance inertial transport, TKE in the bluff-body wake, and pore-scale flow restructuring within the porous layer. Four Reynolds numbers, $Re = 500$, 1000, 2000, and 4000, are examined. As $Re$ increases, the bluff-body wake carries stronger velocity fluctuations toward the impingement region, while pore-scale advection, shear, and obstacle-generated fluctuations also intensify throughout the porous matrix. Consistent with this increase in inertial transport in the impingement and non-impingement regions, the corresponding volume-averaged Nusselt number increases from approximately 20 at $Re = 500$ to approximately 90–100 at $Re = 4000$ in both sampling regions (Fig. 10). The relative interface enhancement is non-monotonic, with $E_{int} = 9.0\%$, 12.7%, 4.8%, and 1.9% for $Re = 500$, 1000, 2000, and 4000, respectively. At higher Reynolds numbers, however, the simultaneous strengthening of pore-scale advection and obstacle-generated fluctuations in both regions raises the non-impingement heat-transfer baseline and reduces the incremental enhancement attributable specifically to direct wake exposure.

The streamwise evolution of $\Delta Nu$ reveals an important distinction between the magnitude of the entrance enhancement and the downstream persistence of the net thermal benefit. At $Re = 500$, $\Delta Nu$ remains positive throughout the sampled porous-layer depth, $7.0 \leq X \leq 8.0$, so no crossover is observed. This behavior is unique among the Reynolds numbers investigated. For $Re = 1000$, 2000, and 4000, $\Delta Nu$ changes sign within the entrance region, with interpolated crossover locations clustered around $X_c \approx 7.35$. Thus, the absence of a crossover at $Re = 500$ demonstrates that the downstream extent of a positive heat-transfer contrast need not coincide with the spatial extent over which wake-specific fluctuation transport remains dynamically important.

The $Re = 500$ case should therefore not be interpreted as evidence that the incident macroscale wake survives throughout the porous layer. The thermal-budget results in Fig. 11 show that the turbulent heat-flux contribution $H_p^t$ is comparatively small at $Re = 500$, while the impingement/non-impingement difference is carried primarily by the mean-convective, dispersive, and wall-input terms. For $Re \geq 1000$, the wake-associated turbulent and dispersive heat flux differences are concentrated near the porous entrance and decay over approximately the same streamwise interval in which $\Delta Nu$ approaches zero at the crossover location $X_c$. The relative mean-velocity deficit $D_U$, however, remains finite beyond this interval.

Reynolds number therefore influences two distinct aspects of the wake-induced thermal response. First, it controls how effectively direct wake impingement is converted into an interface heat-transfer enhancement, with the largest relative gain occurring at $Re = 1000$. Second, it controls the downstream balance between the wake-induced transport enhancement and the competing heat-transfer response of the surrounding porous matrix. Although $D_U$ decreases with increasing Reynolds number, the thermal contrast becomes negative downstream for $Re \geq 1000$ because the favorable wake-associated transport contributions decay while the non-impingement heat-transfer baseline remains comparatively strong. In contrast, at $Re = 500$, a positive $\Delta Nu$ persists throughout the sampled layer despite the comparatively weak turbulent redistribution contribution. This distinction demonstrates that wake-transport persistence and net heat-transfer-enhancement persistence are separate measures.

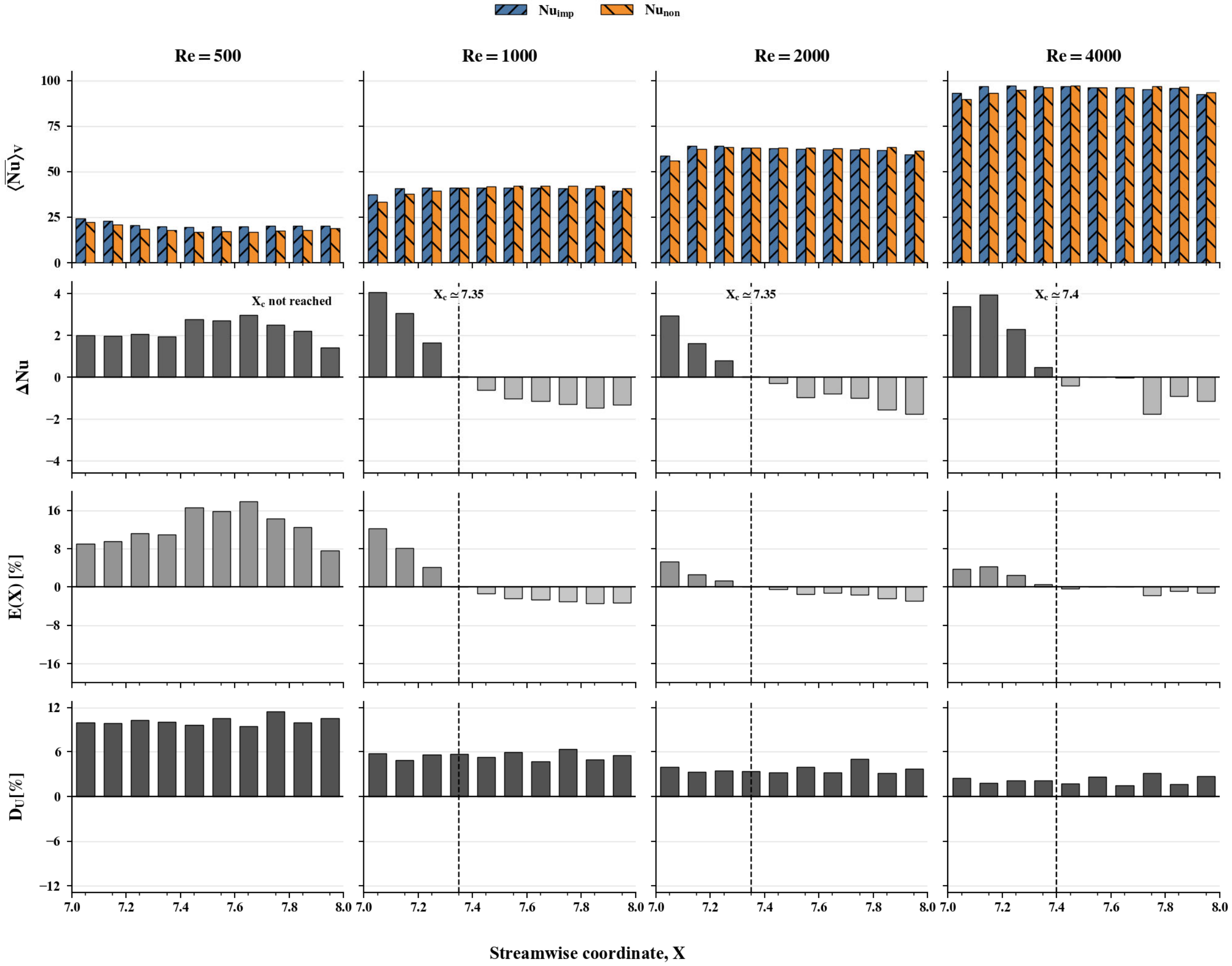

Fig. 10: Streamwise evolution of the region-averaged local-temperature-based Nusselt number (first row), the absolute impingement contrast $\Delta Nu(X)$(second row), the relative contrast $E(X)$(third row), and the local relative mean-velocity deficit $D_U(X)$(fourth row) for $Re = 500$, 1000, 2000, and 4000 at $\phi = 0.75$ and $Pr = 7.0$. The interface enhancements are 9.0%, 12.7%, 4.8%, and 1.9%, respectively. At $Re = 500$, $\Delta Nu$ remains positive through $X = 8.0$, and $X_c$ is therefore not reached within the sampled porous-layer depth. At $Re \geq 1000$, $\Delta Nu$ changes sign between the sampled stations at $X = 7.3$ and 7.4, giving $X_c \approx 7.35$. $D_U$ decreases as $Re$ increases, showing that the downstream thermal penalty is not controlled by an increasing relative mean-velocity deficit.

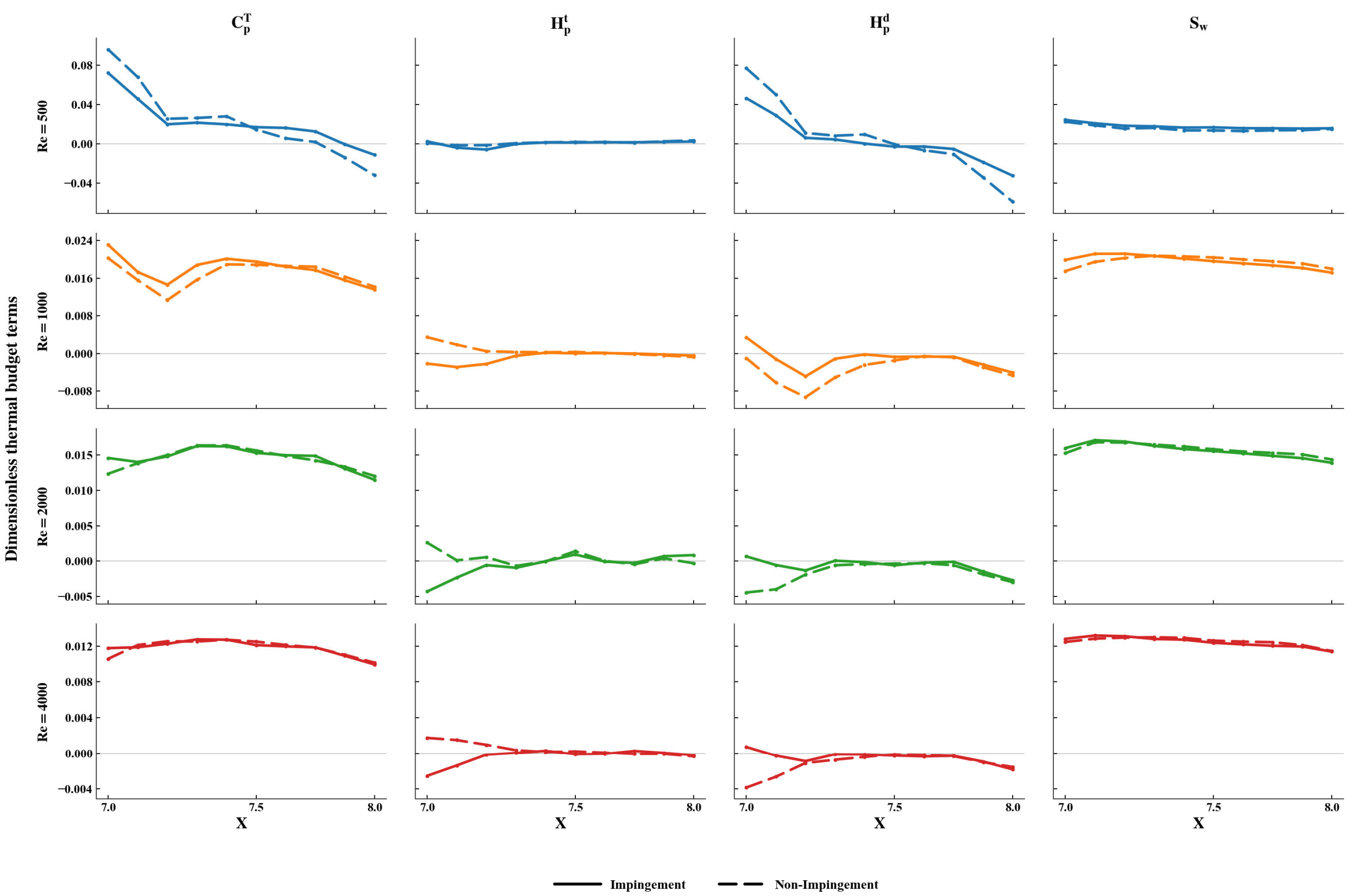

Fig. 11: Thermal-budget terms for $Re$ = 500, 1000, 2000, and 4000 at $\phi$ = 0.75 and $Pr$ = 7.0. Each rows corresponds to a Reynolds number, and the columns show the mean-convective term $C_p^T$, turbulent redistribution $H_p^t$, dispersive redistribution $H_p^d$, and wall input $S_W$. Solid and dashed lines denote the impingement and non-impingement regions, respectively. Within each Reynolds-number row, all terms use the same ordinate scale, while the scale varies between rows. At $Re$ = 500, $H_p^t$ is small, and the positive Nusselt-number contrast is associated mainly with differences in $C_p^T$, $H_p^d$, and $S_W$. At $Re \geq 1000$, the impingement/non-impingement differences in $H_p^t$ and $H_p^d$ are concentrated near the entrance and weaken over the interval in which $\Delta Nu$ changes sign.

The Prandtl-number comparison at $Re = 1000$ and $\phi = 0.75$ isolates the influence of thermal diffusivity while maintaining the same momentum forcing. Figure 12 shows that the absolute local-temperature-based Nusselt number level is higher at $Pr$ = 7.0 than at $Pr$ = 0.7.

At the porous-fluid interface, however, the lower-Prandtl-number case exhibits the greater sensitivity to direct wake exposure. The interface enhancement is $E_{int} \approx 16\%$ for Pr = 0.7, compared with 12.7% for $Pr = 7.0$. In contrast, the absolute Nusselt-number difference $\Delta Nu$ is larger at $Pr = 7.0$ because the baseline Nusselt-number scale is higher. The two metrics therefore provide complementary information: $\Delta Nu$ measures the absolute separation between the impingement and non-impingement regions, whereas $E_{int}$ measures the wake-induced gain relative to the local non-impingement baseline.

Despite the difference in entrance amplitude, the downstream evolution is similar for the two Prandtl numbers. For both $Pr = 0.7$ and $Pr = 7.0$, $\Delta Nu$ changes sign within the same narrow entrance interval, with interpolated values clustered around $X_c \approx 7.35$ (Fig. 12). Within the present streamwise resolution, no measurable displacement of the crossover location with Prandtl number is observed. Thus, the larger relative entrance enhancement at $Pr = 0.7$ does not translate into a longer region of positive wake-induced heat-transfer contrast. Farther downstream, both cases exhibit $\Delta Nu < 0$. The absolute deficit is larger at $Pr = 7.0$, consistent with its larger overall Nusselt-number scale, while the relative downstream penalties remain of comparable magnitude.

For both fluids, the wake-associated differences in the turbulent and dispersive contributions are concentrated near the porous entrance and decay over approximately the same streamwise interval in which $\Delta Nu$ approaches zero (Fig. 13). Once these wake-specific contributions weaken, the remaining mean-flow disadvantage is sufficient to produce the downstream negative $\Delta Nu$. The results therefore indicate that Prandtl number modifies primarily the magnitude of the wake-induced thermal response, while the streamwise extent of the favorable response remains essentially unchanged within the present resolution.

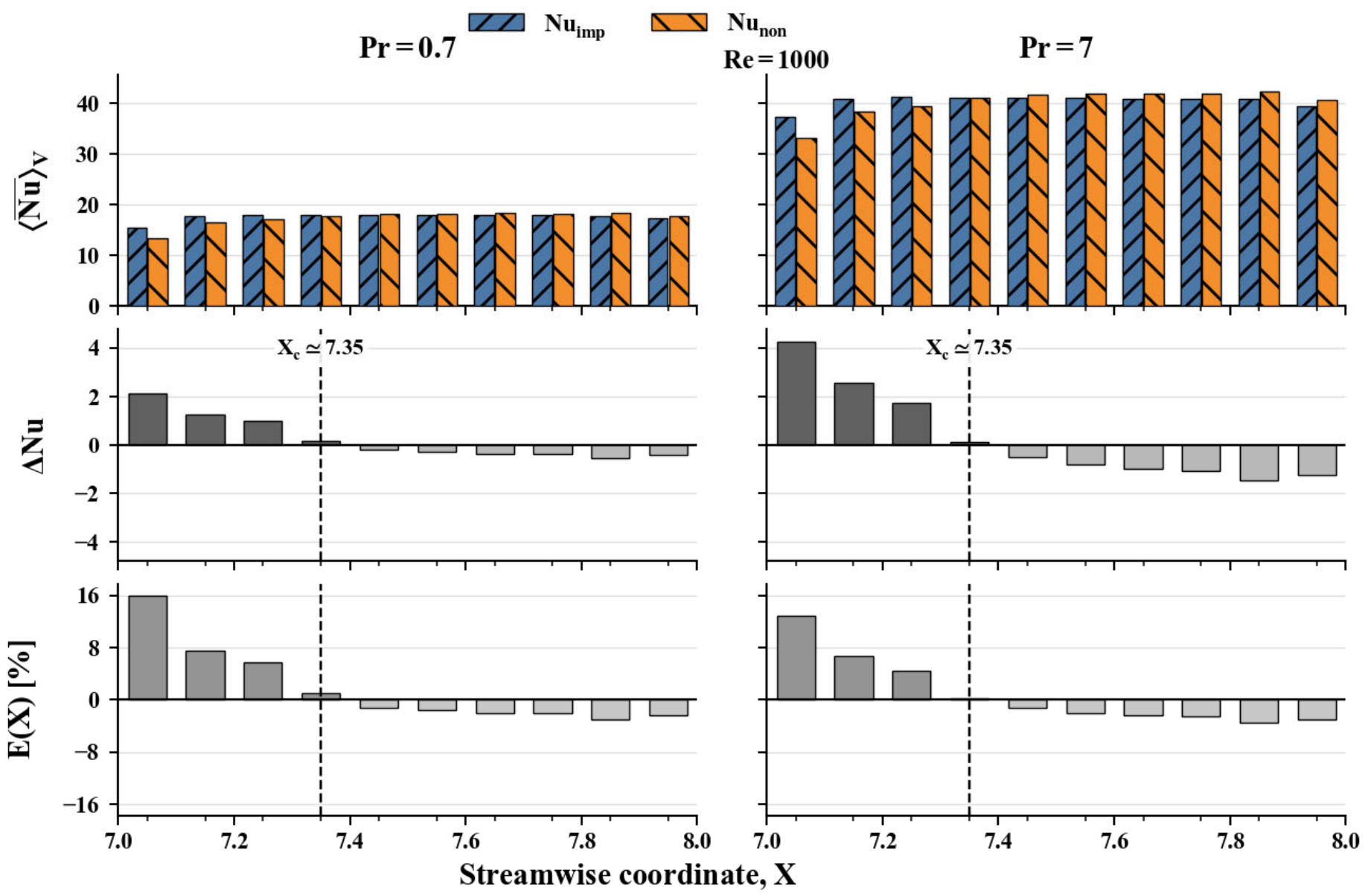

Fig. 12: Effect of Prandtl number on the wake-specific thermal response at $Re$ = 1000 and $\phi$ = 0.75. The rows show $\langle \overline{Nu} \rangle_V$, $\Delta Nu(X)$, and $E(X)$ for $Pr$ = 0.7 and $Pr$ = 7.0. Both cases exhibit a positive entrance contrast followed by a sign reversal near $X_c \approx 7.35$. Increasing Pr raises the absolute Nusselt-number level and absolute entrance contrast but produces little measurable change in the crossover position.

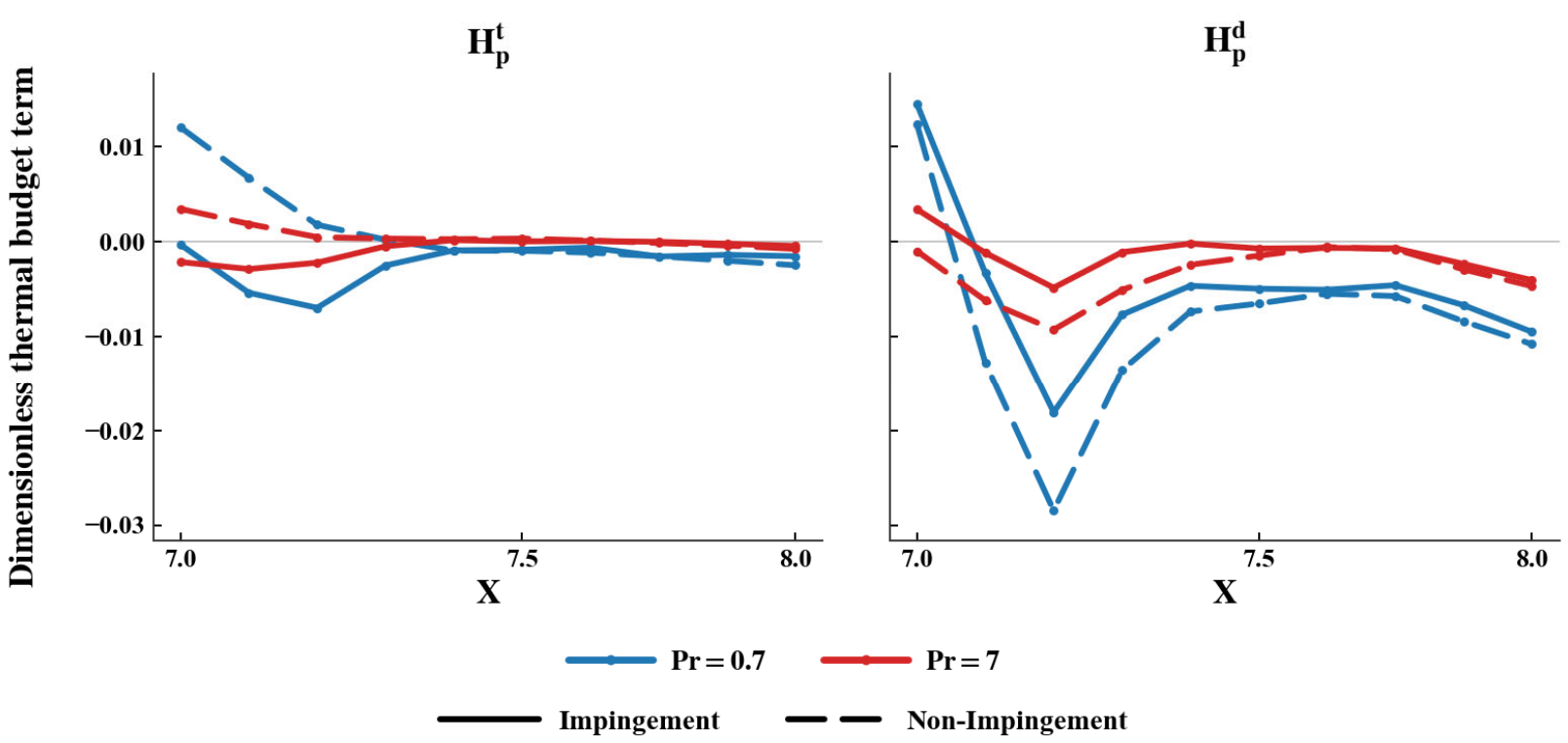

Fig. 13: Streamwise variation of the volume-averaged thermal-budget terms in the impingement and non-impingement regions for $Pr = 0.7$ and $Pr = 7.0$ at fixed $Re = 1000$ and $\phi = 0.75$. The panels show the turbulent heat-flux redistribution $H_p^t$ and dispersive heat-flux redistribution $H_p^d$. Solid and dashed lines denote the impingement and non-impingement regions, respectively. The lower-Prandtl-number case exhibits stronger impingement/non-impingement differences in $H_p^t$ and $H_p^d$ near the porous entrance. These differences decrease rapidly downstream for both Prandtl numbers, showing that thermal diffusivity changes the strength of the wake-induced thermal redistribution more than the streamwise extent over which the impingement region remains distinct from the non-impingement baseline.

## 4 Conclusions

Two-dimensional pore-resolved simulations reveal a two-stage thermal response to square-bluff-body wake impingement on a heated porous layer. Direct wake impingement initially enhances heat transfer at the first heated obstacles, producing higher surface-averaged wall heat flux and local-temperature-based Nusselt number relative to a geometrically matched non-impingement region. As the flow penetrates the porous matrix, the imposed fluctuation field is progressively reorganized by dispersion within the porous matrix. TKE is produced inside the matrix at the microscale level by the solid obstacles independently of the introduction of TKE by the bluff-body wake. Consistent with this interpretation, the no-bluff-body inlet-condition controls show that substantial pore-scale fluctuation energy can develop even from a laminar inlet, whereas the additional TKE, Nusselt-number response, and wall heat flux associated with the synthetic-vortex inlet progressively diminish relative to the laminar-inlet case.

The thermal budget analysis further supports this interpretation. Near the porous-fluid interface, direct impingement produces additional turbulent and dispersive thermal transport relative to the non-impingement region. Farther downstream, the wake-associated turbulent contribution becomes small, while the impingement/non-impingement difference in dispersive transport relaxes toward the pore-generated background. Mean convection, wall heat input, and finite pore-scale fluctuations remain present, indicating that heat transfer continues after the additional macroscale-to-microscale redistribution associated with the incoming wake has been completed.

The primary result of this study is the relatively consistent length of the wake-associated macroscale thermal entrance region. Across all investigated porosities, Reynolds numbers, and Prandtl numbers, the macroscale turbulent heat-flux contribution decays within approximately 3–4 unit cells. We believe this parameter-insensitive length represents a fundamental macroscale thermal entrance scale for the present class of porous-fluid interactions: it marks the conversion of externally generated vortex impingement into predominantly local pore-scale thermal transport.

However, the magnitude of entrance heat transfer enhancement is not universal. Porosity produces a non-monotonic relative response, with the largest interface enhancement ($E_{int} = 18.2\%$) occurring at $\phi$ = 0.85. Reynolds number also produces a non-monotonic response, with $E_{int}$ = 12.7% for $Re = 1000$. The Prandtl-number comparison further shows that $Pr = 0.7$ produces the larger relative interface enhancement, whereas $Pr = 7.0$ produces the larger absolute Nusselt-number difference. These results show that pore geometry, inertia, and thermal diffusivity strongly modify the magnitude of the wake-induced thermal response, even though there is no measurable change in the observed 3–4 unit-cell wake-associated entrance length.

Importantly, this macroscale thermal entrance scale is distinct from the downstream distance over which the net Nusselt-number contrast remains positive. For the $Re \geq 1000$ configurations investigated here, including the porosity and Prandtl-number comparisons performed at $Re = 1000$, $\Delta Nu$ changes sign near $X_c \approx 7.35$. In contrast, at $Re = 500$, $\Delta Nu$ remains positive throughout the sampled depth up to $X = 8.0$ even though the wake-associated turbulent contribution is comparatively small farther downstream. The $Re = 500$ case therefore demonstrates that decay of the additional macroscale turbulent transport contribution does not necessarily coincide with the disappearance of the net heat-transfer benefit.

These results also suggest a bounded design implication. For the present configurations, maximizing $E_{int}$ does not necessarily maximize the absolute Nusselt number or extend the downstream region of positive $\Delta Nu$. For the cases $Re \geq 1000$, the favorable wake-induced contrast is confined primarily to the entrance region, after which the impingement region can develop a heat-transfer penalty. This behavior suggests that thermal loading concentrated near the wake-impingement entrance region may make more effective use of a single upstream wake source than loading distributed uniformly farther downstream. Sustained enhancement at greater depths would likely require additional or distributed mixing rather than reliance on a single upstream disturbance. This implication remains specific to the present two-dimensional geometry and should be evaluated in three-dimensional and geometrically varied porous configurations before being used as a general design recommendation.

Future work should therefore test the effects of the bluff-body-to-pore-scale ratio, pore arrangement, porosity distribution, and porous-layer depth. In addition, the incident wake should be characterized independently through quantities such as TKE, mean-velocity deficit, wake width, dominant shedding frequency, integral length scale, spatial correlations, and spectral content. Such comparisons would establish whether the wake-associated thermal entrance scale of approximately 3–4 unit cells observed here remains robust when the wake structure and porous geometry are varied independently.

**Acknowledgements.** The authors acknowledge the computing resources provided by North Carolina State University High Performance Computing Services Core Facility (RRID:SCR_022168). AVK acknowledges the support of the Alexander von Humboldt Foundation through the Humboldt Research Award.
**Funding.** This research was funded by the National Science Foundation under award CBET-2042834.
**Declarations of interests.** The authors report no conflict of interest.
**Data availability statement.** Data sharing is not applicable – no new data or custom code were generated for this manuscript. ANSYS Fluent 21.2 was used to solve the governing equations.

**References**

[1] R. Rajagopalan, C. Tien, Trajectory analysis of deep-bed filtration with the sphere-in-cell porous media model, AIChE Journal 22 (1976) 523–533. https://doi.org/10.1002/aic.690220316.
[2] N.S. Hanspal, V. Nassehi, A. Kulkarni, Three-dimensional finite element modelling of coupled free/porous flows: applications to industrial and environmental flows, Int. J. Numer. Methods Fluids 71 (2013) 1382–1421. https://doi.org/10.1002/fld.3717.
[3] W. Mell, A. Maranghides, R. McDermott, S.L. Manzello, Numerical simulation and experiments of burning douglas fir trees, Combust. Flame 156 (2009) 2023–2041. https://doi.org/10.1016/j.combustflame.2009.06.015.
[4] M. Zamani, A. Sangtarash, M. Javad Maghrebi, Numerical Study of Porous Media Effect on the Blade Surface of Vertical Axis Wind Turbine for Enhancement of Aerodynamic Performance, Energy Convers. Manag. 245 (2021) 114598. https://doi.org/10.1016/j.enconman.2021.114598.
[5] M.E. Rosti, L. Brandt, A. Pinelli, Turbulent channel flow over an anisotropic porous wall – drag increase and reduction, J. Fluid Mech. 842 (2018) 381–394. https://doi.org/10.1017/jfm.2018.152.
[6] N. Abderrahaman-Elena, R. García-Mayoral, Analysis of anisotropically permeable surfaces for turbulent drag reduction, Phys. Rev. Fluids 2 (2017) 114609. https://doi.org/10.1103/PhysRevFluids.2.114609.
[7] M. Itoh, S. Tamano, R. Iguchi, K. Yokota, N. Akino, R. Hino, S. Kubo, Turbulent drag reduction by the seal fur surface, Physics of Fluids 18 (2006) 065102. https://doi.org/10.1063/1.2204849.
[8] C.Y. Zhao, T.J. Lu, Analysis of microchannel heat sinks for electronics cooling, Int. J. Heat Mass Transf. 45 (2002) 4857–4869. https://doi.org/10.1016/S0017-9310(02)00180-1.
[9] C.-W. Huang, V. Srikanth, A. V. Kuznetsov, The evolution of turbulent micro-vortices and their effect on convection heat transfer in porous media, J. Fluid Mech. 942 (2022) A16. https://doi.org/10.1017/jfm.2022.291.
[10] X. Chu, G. Yang, S. Pandey, B. Weigand, Direct numerical simulation of convective heat transfer in porous media, Int. J. Heat Mass Transf. 133 (2019) 11–20. https://doi.org/10.1016/j.ijheatmasstransfer.2018.11.172.
[11] L. Chen, Y. Zhao, Y. Gao, B. Weigand, Detailed numerical study on jet impingement heat transfer using porous lattice structures and multi-objective optimization of lattice cells, Appl. Therm. Eng. 279 (2025) 127705. https://doi.org/10.1016/j.applthermaleng.2025.127705.
[12] B. Mayer, Investigations of Pressure Loss and Heat Transfer in a Regular Metallic Porous Structure, Ph.D. dissertation, University of Stuttgart, 2014.

[13] K. Suga, Y. Matsumura, Y. Ashitaka, S. Tominaga, M. Kaneda, Effects of wall permeability on turbulence, Int. J. Heat Fluid Flow 31 (2010) 974–984. https://doi.org/10.1016/j.ijheatfluidflow.2010.02.023.
[14] C. Manes, D. Poggi, L. Ridolfi, Turbulent boundary layers over permeable walls: scaling and near-wall structure, J. Fluid Mech. 687 (2011) 141–170. https://doi.org/10.1017/jfm.2011.329.
[15] Y. Kuwata, K. Suga, Extensive investigation of the influence of wall permeability on turbulence, Int. J. Heat Fluid Flow 80 (2019) 108465. https://doi.org/10.1016/j.ijheatfluidflow.2019.108465.
[16] Z. Hao, R. García-Mayoral, Turbulent flows over porous and rough substrates, J. Fluid Mech. 1008 (2025) A1. https://doi.org/10.1017/jfm.2025.55.
[17] W. Wang, X. Chu, A. Lozano-Durán, R. Helmig, B. Weigand, Information transfer between turbulent boundary layers and porous media, J. Fluid Mech. 920 (2021) A21. https://doi.org/10.1017/jfm.2021.445.
[18] J. Courter, V. Srikanth, T. Kemayo, A. V. Kuznetsov, Scale Collapse of Vortices at Porous-Fluid Interfaces, (2026). https://doi.org/10.48550/arXiv.2601.10396.

# Macroscale vortex impingement at a porous-fluid interface induces local heat-transfer enhancement

Thibaut K. Kemayo, Vishal Srikanth, Justin Courter, Rodrigo R. Caballero, and Andrey V. Kuznetsov
Department of Mechanical and Aerospace Engineering, North Carolina State University, Raleigh, NC 27695, USA
*Corresponding author: avkuznet@ncsu.edu*

# Supplementary Material

## S1. No-bluff-body inlet-condition controls

### S1.1 Control configuration and synthetic-vortex inlet specification

To determine whether substantial downstream velocity fluctuation energy can develop from pore-scale dynamics in the absence of a bluff-body wake, two no-bluff-body control simulations were performed at $Re = 1000$, $Pr = 7.0$, and $\phi = 0.75$. The porous geometry, heated-obstacle boundary condition, inlet temperature, numerical procedure, time-averaging procedure, and pore-control-volume sampling were kept the same as in the reference configuration. The two controls therefore differ only in the imposed inlet condition.

The laminar-inlet control used a uniform streamwise inlet velocity. The second control used a synthetic-vortex velocity inlet generated with an ANSYS Fluent user-defined function (UDF). A continuous train of compact two-dimensional vortices was superimposed on a mean inlet velocity $u_{in} = 1\ m\ s^{-1}$. The 60-m-high inlet contained six transverse vortex centers at y = 5, 15, 25, 35, 45, and 55 m. Each vortex had a diameter of 10 m (radius R = 5 m). Successive vortex columns were separated by 8 m and convected at 1 m $s^{-1}$, corresponding to an injection period of 8 s. All imposed vortices rotated in the same direction.

The vortex amplitude was selected to achieve a target two-dimensional TKE of 0.3 $m^2\ s^{-2}$ so that the turbulent-energy magnitude at the first sampled porous station was comparable to that of the bluff-body impingement region in the reference case. The prescribed dimensional target corresponds to a nondimensional value of 0.3 when normalized by $u_{in}^2$; however, the porous-entrance value is evaluated after the imposed vortices interact with the first porous obstacles and is therefore assessed separately from the prescribed inlet target. This control is not intended to reproduce broadband turbulence or the bluff-body wake. In particular, it does not reproduce the wake spatial distribution, mean-velocity deficit, characteristic shedding scale, integral length scale, spectral content, or detailed vortical organization.

### S1.2 Comparison of metrics

The streamwise curves and differences reported in Figs. S3–S5 are evaluated using the same pore-control-volume sampling procedure for both no-bluff-body control cases. The same procedure is used for the porous-entrance TKE comparison at the first sampled pore-control volume, $X_0 = 7.05$. Because the sampling geometry is identical for the laminar-inlet and synthetic-vortex inlet controls, the resulting differences isolate the effect of the imposed inlet condition rather than differences in spatial averaging.

Using “syn” for the synthetic-vortex inlet control and “lam” for the laminar-inlet control, the synthetic-minus-laminar differences in turbulent energy and thermal response are defined as follows. The subscript IT denotes the inlet-condition contrast between the synthetic-vortex and laminar controls. The TKE and Nusselt-number definitions are identical to those used in the main paper. The dimensionless TKE is normalized by $u_{in}^2$, while the local-temperature-based mean Nusselt number uses the intrinsic volume-averaged fluid temperature in the thermal driving difference. The resulting quantities therefore measure the additional fluctuation and thermal response associated with the synthetic-vortex inlet relative to the laminar-inlet control at corresponding streamwise pore-control-volume locations.

$$\Delta\langle \mathrm{TKE}\rangle_{\mathrm{IT}}(\mathrm{X}) = \langle \mathrm{TKE}\rangle_{\mathrm{V,syn}}(\mathrm{X}) - \langle \mathrm{TKE}\rangle_{\mathrm{V,lam}}(\mathrm{X}) \qquad S1$$

$$\Delta Nu_{\mathrm{IT}}(\mathrm{X}) = \langle \overline{Nu}\rangle_{\mathrm{V,syn}}(\mathrm{X}) - \langle \overline{Nu}\rangle_{\mathrm{V,lam}}(\mathrm{X}) \qquad S2$$

$$\Delta\langle \mathrm{q}''\rangle_{\mathrm{IT}}(\mathrm{X}) = \left\langle \overline{\mathrm{q}''}\right\rangle_{\mathrm{S,syn}}(\mathrm{X}) - \langle \overline{\mathrm{q}''}\rangle_{\mathrm{S,lam}}(\mathrm{X}) \qquad S3$$

$$E_{\mathrm{IT}}(\mathrm{X}) = \frac{\Delta Nu_{\mathrm{IT}}(\mathrm{X})}{\langle \overline{Nu}\rangle_{\mathrm{V,lam}}(\mathrm{X})} \times 100\% \qquad S4$$

Positive values indicate a larger response for the synthetic-vortex inlet. To compare downstream attenuation independently of the entrance difference, the normalized contrasts are defined using the first sampled pore-control volume, $X_0 = 7.05$. Because the denominator is the signed difference at $X_0$, values above unity indicate local amplification relative to the first sampled station, values approaching zero indicate loss of the inlet-condition contrast, and negative values indicate a local reversal of the synthetic-minus-laminar difference.

$$\mathrm{M_K}(\mathrm{X}) = \frac{\Delta\langle TKE\rangle_{IT}(X)}{\Delta\langle TKE\rangle_{IT}(X0)} \qquad S5$$

$$M_{\mathrm{Nu}}(\mathrm{X}) = \frac{\Delta Nu_{IT}(X)}{\Delta Nu_{IT}(X0)} \qquad S6$$

$$M_{\mathrm{q}}(\mathrm{X}) = \frac{\Delta\langle q''\rangle_{IT}(X)}{\Delta\langle q''\rangle_{IT}(X0)} \qquad S7$$

The normalization is used only as a compact measure of how the difference between the two inlet conditions evolves with X; it is not interpreted as a decay law for the bluff-body wake.

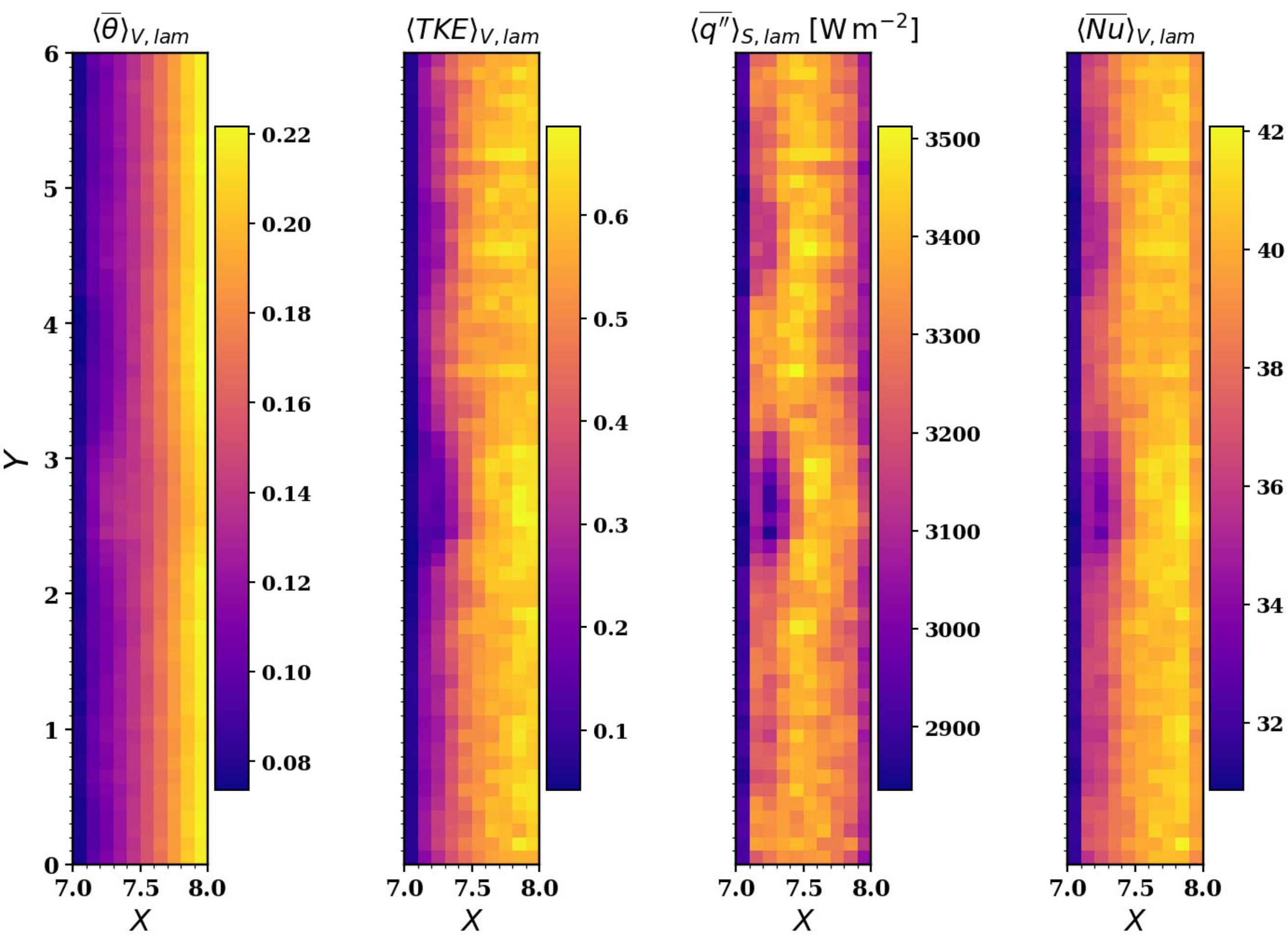

Fig. S1. Spatial distributions of (from left to right) volume-averaged mean dimensionless fluid temperature, volume-averaged turbulent kinetic energy, surface-averaged wall heat flux, and local-temperature-based mean Nusselt number for the no-bluff-body laminar-inlet control at $\boldsymbol{Re = 1000}$, $\boldsymbol{Pr = 7.0}$, and $\boldsymbol{\phi = 0.75}$ over $\boldsymbol{7.0 \le X \le 8.0}$. The corresponding streamwise comparisons are shown in Figs. S3–S5. The laminar inlet supplies little turbulent energy at the porous entrance, whereas substantial TKE develops farther inside the matrix as the flow undergoes repeated pore-throat acceleration, shear, separation, and obstacle-wake formation. This case provides the pore-generated baseline for the inlet-condition comparisons.

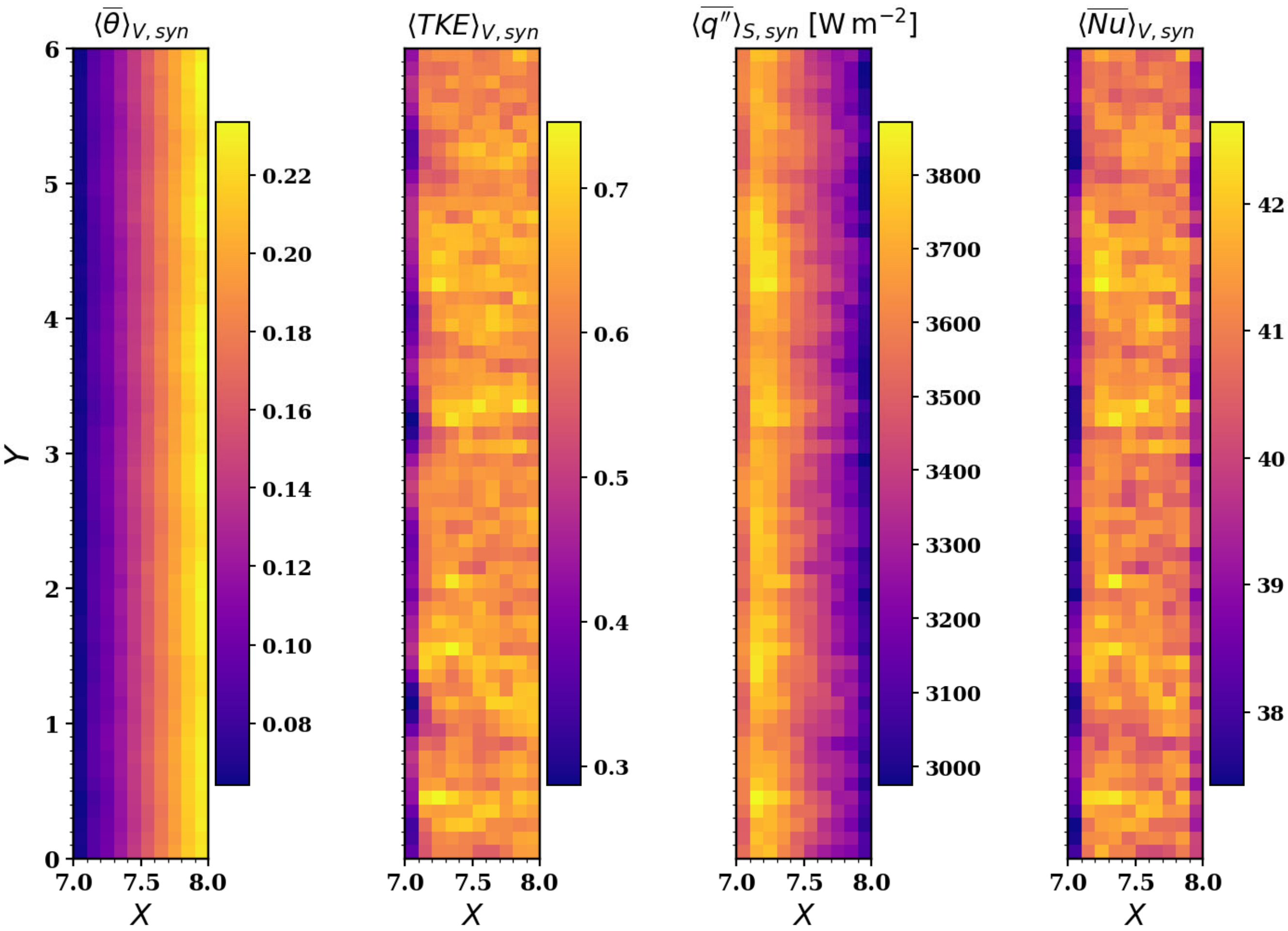

Fig. S2. Spatial distributions of (from left to right) volume-averaged mean dimensionless fluid temperature, volume-averaged turbulent kinetic energy, surface-averaged wall heat flux, and local-temperature-based mean Nusselt number for the no-bluff-body synthetic-vortex inlet control at $\boldsymbol{Re = 1000}$, $\boldsymbol{Pr = 7.0}$, and $\boldsymbol{\phi = 0.75}$ over $\boldsymbol{7.0 \le X \le 8.0}$. The corresponding streamwise comparisons are shown in Figs. S3–S5. The fields show how the matrix modifies the imposed fluctuation field through repeated pore-scale acceleration, obstruction, shear, and separation.

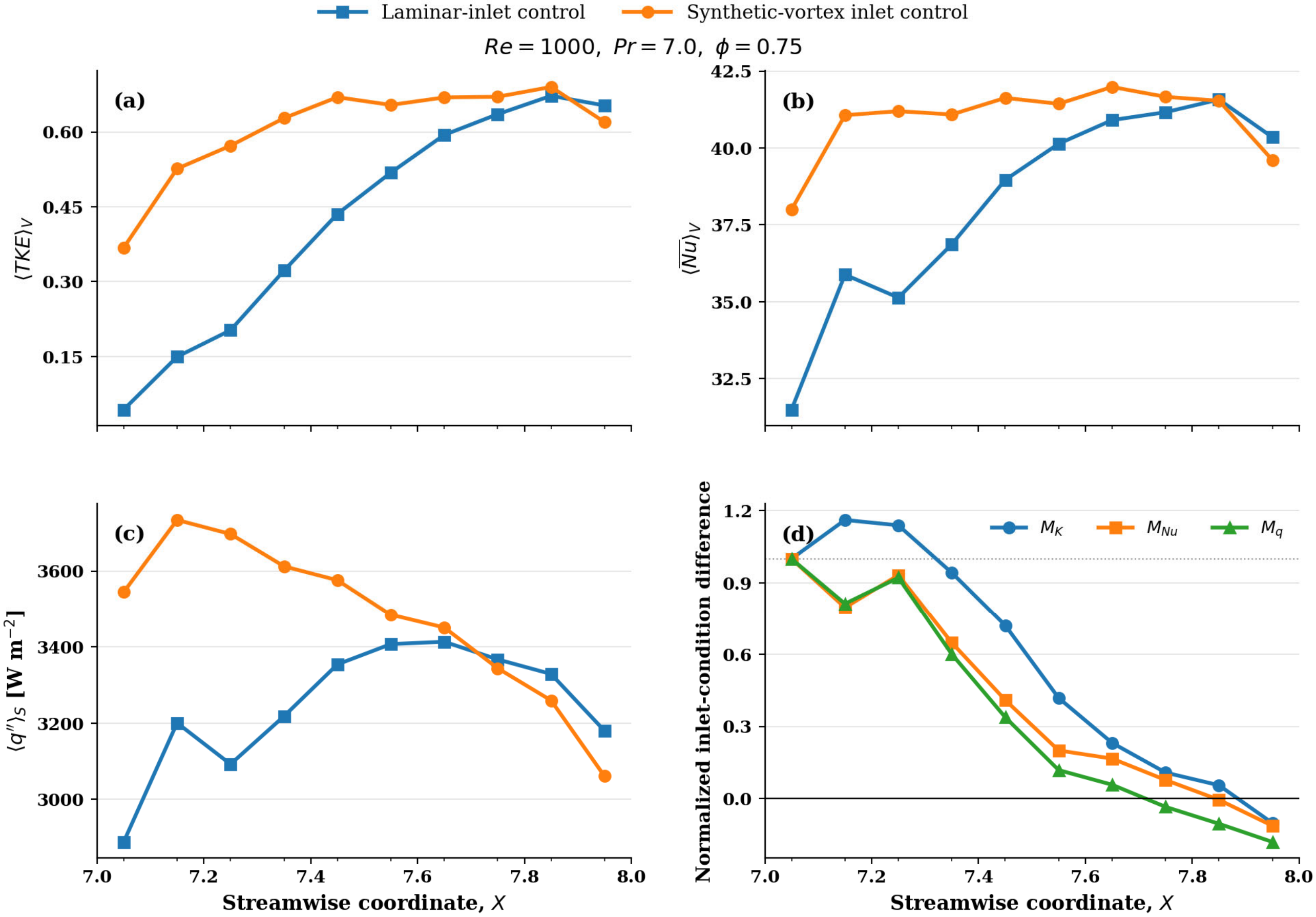

Fig. S3. Streamwise evolution of the no-bluff-body laminar-inlet and synthetic-vortex inlet controls at $\boldsymbol{Re} = \mathbf{1000}$, $\boldsymbol{Pr} = \mathbf{7.0}$, and $\boldsymbol{\phi} = \mathbf{0.75}$. Panels show (a) volume-averaged turbulent kinetic energy $\langle \boldsymbol{TKE} \rangle_{\boldsymbol{V}}$, (b) local-temperature-based mean Nusselt number $\langle \overline{\boldsymbol{Nu}} \rangle_{\boldsymbol{V}}$, (c) surface-averaged wall heat flux $\langle \boldsymbol{q''} \rangle_{\boldsymbol{S}}$, and (d) the normalized synthetic-minus-laminar inlet-condition differences $\boldsymbol{M_K}$, $\boldsymbol{M_{Nu}}$, and $\boldsymbol{M_q}$, each normalized by its value in the first sampled pore-control volume at $\boldsymbol{X_0} = \mathbf{7.05}$. The laminar-inlet control begins with substantially lower turbulent energy and thermal transport near the porous entrance, while pore-generated TKE increases rapidly downstream. Over the same streamwise interval, the differences in Nusselt number and wall heat flux between the two inlet conditions decrease and become small near the downstream end of the sampled layer. In panel (d), $\boldsymbol{M_K}$ initially increases above unity, indicating a short downstream amplification of the $\boldsymbol{TKE}$ difference before its subsequent decay, whereas $\boldsymbol{M_{Nu}}$ and $\boldsymbol{M_q}$ exhibit smaller non-monotonic variations followed by an overall reduction toward zero. Negative normalized values indicate locations where the laminar-inlet response exceeds the synthetic-vortex inlet response. The combined behavior therefore indicates an overall downstream loss of inlet-condition contrast rather than a monotonic decay.

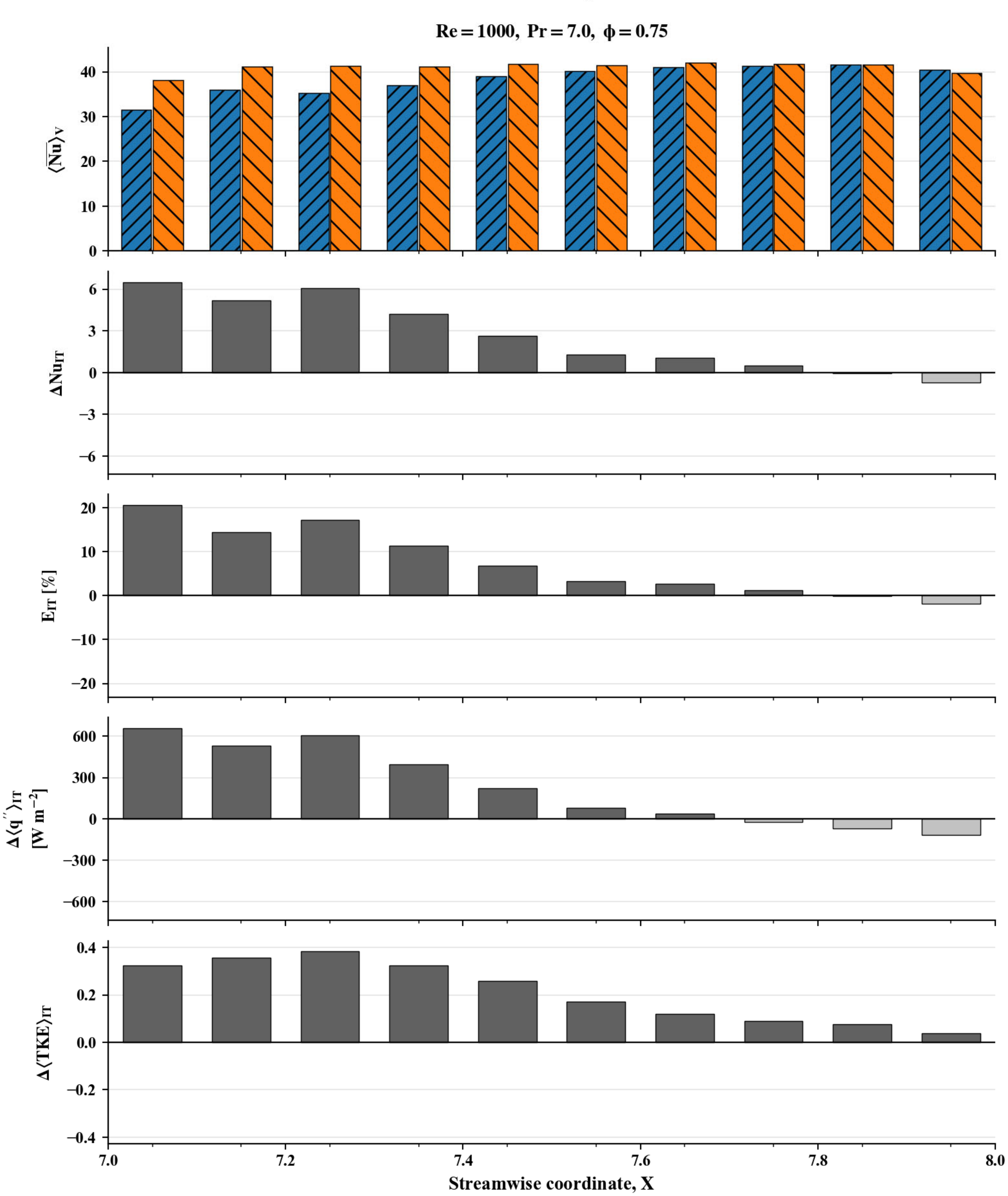

Fig. S4. Signed streamwise summary of the laminar-inlet and synthetic-vortex inlet controls at $\boldsymbol{Re = 1000}$, $\boldsymbol{Pr = 7.0}$, and $\boldsymbol{\phi = 0.75}$. The first row shows the local-temperature-based mean Nusselt number for both inlet conditions. The remaining rows show the synthetic-minus-laminar increments $\boldsymbol{\Delta Nu_{IT}}$, $\boldsymbol{E_{IT}}$, $\boldsymbol{\Delta\langle q''\rangle_{IT}}$, and $\boldsymbol{\Delta\langle TKE\rangle_{IT}}$. The additional turbulent and thermal responses are largest near the entrance and decrease downstream; the final negative bars indicate local reversal rather than disappearance of all pore-scale fluctuations. This figure provides the signed magnitudes that complement the raw curves in Fig. S3(a–c) and the normalized evolution in Fig. S3(d).

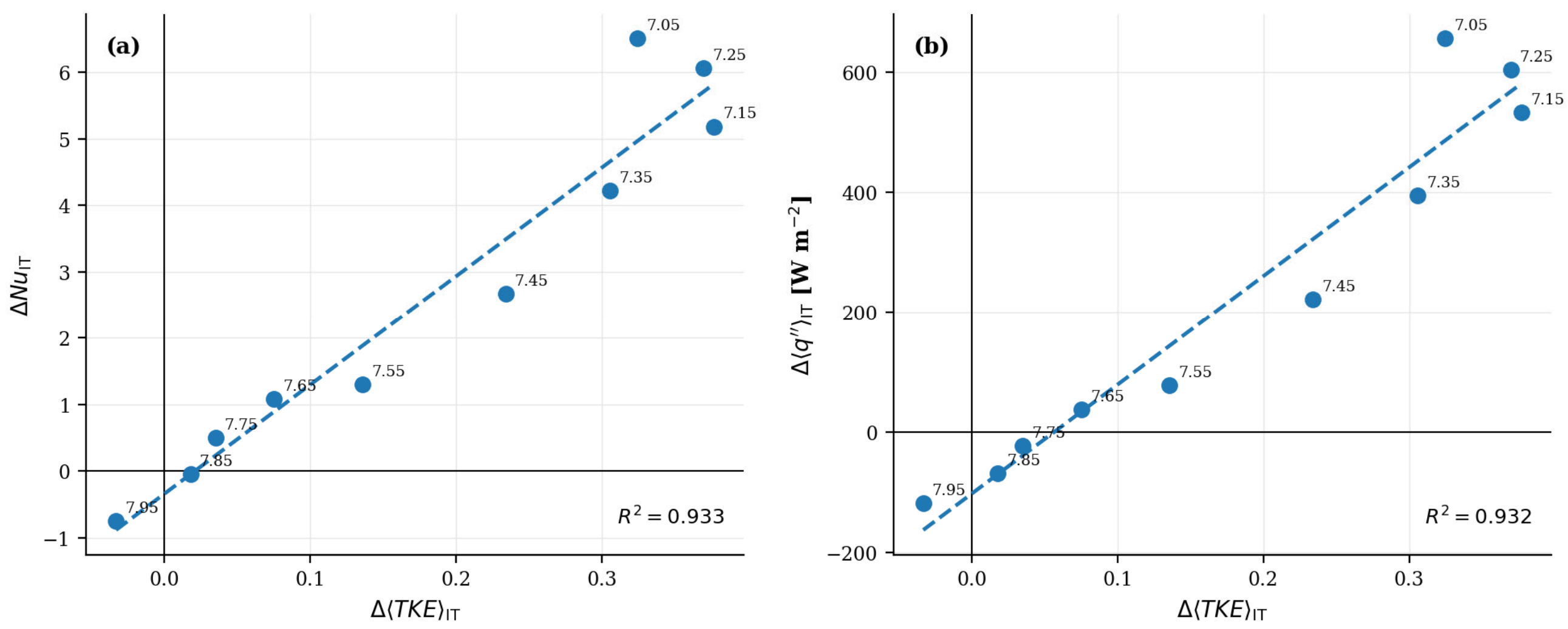

Fig. S5. Relation between the inlet-condition TKE difference and the corresponding thermal differences for the no-bluff-body controls at $\boldsymbol{Re = 1000}$, $\boldsymbol{Pr = 7.0}$, and $\boldsymbol{\phi = 0.75}$: (a) $\boldsymbol{\Delta Nu_{IT}}$ versus $\boldsymbol{\Delta\langle TKE\rangle_{IT}}$ and (b) $\boldsymbol{\Delta\langle q''\rangle_{IT}}$ versus $\boldsymbol{\Delta\langle TKE\rangle_{IT}}$. Each symbol represents one of the 10 successive streamwise pore-control-volume stations and is labeled by its X coordinate; dashed lines denote least-squares linear fits. The fitted $R^2$ values are 0.933 and 0.932, respectively. Because both axes vary systematically with the same streamwise coordinate and the stations are spatially ordered rather than independent replicate samples, these $R^2$ values are interpreted only as descriptive streamwise associations. They do not establish a causal one-to-one relation between turbulent energy and thermal response and are not used as evidence of bluff-body wake persistence.

## S1.3 Interpretation of control cases

Figures S1–S4 show that the inlet-condition contrast is large near the porous entrance and is progressively reduced as the flow develops through the matrix. The laminar-inlet control begins with weak turbulent energy but develops substantial downstream TKE, while the Nusselt-number and wall-heat-flux differences relative to the synthetic-vortex inlet decrease toward zero and become slightly negative near the end of the sampled layer. This behavior is consistent with the main-paper observation that finite downstream TKE can remain substantial even after the wake-specific Nusselt-number advantage has disappeared, because pore-throat acceleration, shear-layer formation, separation, and repeated obstacle wakes continuously generate local fluctuations. The controls therefore demonstrate that downstream TKE magnitude alone cannot identify persistence of an externally generated wake.

Figure S5 further shows that the synthetic-minus-laminar thermal differences vary in the same overall streamwise direction as the turbulent-energy difference, but the streamwise regressions are treated only as descriptive associations. The synthetic-vortex inlet is not a surrogate for the bluff-body wake: only its porous-entrance turbulent-energy magnitude was tuned to be comparable, whereas the reference wake also contains a localized mean-velocity deficit and a distinct spatial and dynamical structure. Accordingly, these controls are not used to determine the wake-specific crossover $Xc$ or the macroscale thermal entrance length of 3–4 unit cells reported in the main paper. Their role is limited to independently supporting the statement that appreciable pore-generated downstream turbulent energy can arise without persistence of the original bluff-body wake.